%% file: paper.tex
\documentclass[preprint,authoryear]{elsarticle}

\usepackage{amssymb}
\usepackage{booktabs}
\usepackage{subcaption}
\usepackage{longtable}
\usepackage{natbib}
\usepackage{hyperref}
\usepackage{amsmath}
\usepackage{lscape}
\usepackage[T1]{fontenc}
\usepackage{amsfonts} 
\usepackage{placeins}
\usepackage{multirow}

\journal{Expert Systems with Applications}
\usepackage{xcolor}

\usepackage{subcaption}
\usepackage{pdflscape}

\begin{document}

\let\WriteBookmarks\relax
\def\floatpagepagefraction{1}
\def\textpagefraction{.001}

\begin{frontmatter}



\title{Forecasting Bitcoin volatility spikes from whale transactions and CryptoQuant data using Synthesizer Transformer models }


\author[label1]{Dorien Herremans\footnote[1]{The authors contributed equally (co-first authors).}}
\ead{dorien_herremans@sutd.edu.sg}
\author[label1]{Kah Wee Low\footnotemark[1]}
\ead{lowkahwee1995@hotmail.com}

\affiliation[label1]{organization={SUTD},
            addressline={Somapah Road 8}, 
            city={Singapore},
            postcode={487372}, 
            state={Singapore},
            country={SG}}



\begin{abstract}

The cryptocurrency market is highly volatile compared to traditional financial markets. Hence, forecasting its volatility is crucial for risk management. In this paper, we investigate CryptoQuant data (e.g. on-chain analytics, exchange and miner data) and whale-alert tweets, and explore their relationship to Bitcoin's next-day volatility, with a focus on extreme volatility spikes. We propose a deep learning Synthesizer Transformer model for forecasting volatility. Our results show that the model outperforms existing state-of-the-art models when forecasting extreme volatility spikes for Bitcoin using CryptoQuant data as well as whale-alert tweets. We analysed our model with the Captum XAI library to investigate which features are most important. We also backtested our prediction results with different baseline trading strategies and the results show that we are able to minimize drawdown while keeping steady profits. Our findings underscore that the proposed method is a useful tool for forecasting extreme volatility movements in the Bitcoin market.

\end{abstract}

\begin{graphicalabstract}
\includegraphics[width=14cm]{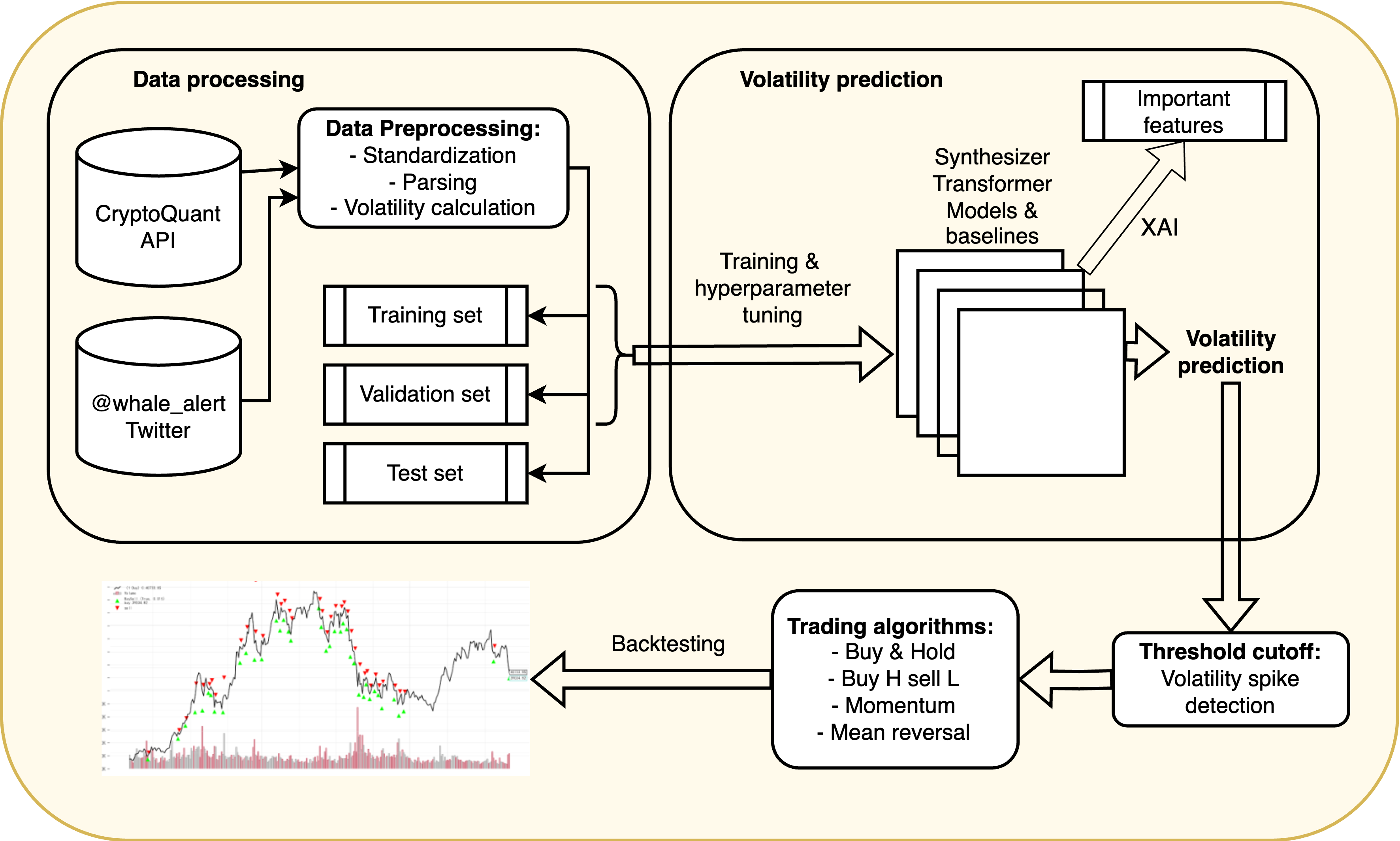}
\end{graphicalabstract}

\begin{highlights}
\item We propose a new Synthesizer Transformer model to predict next-day BTC volatility.
\item Our predictive model takes as input whale-alert data from Twitter.
\item Our model also uses CryptoQuant data, which includes on-chain and exchange data.
\item Explainable AI techniques (XAI) are used to uncover important features. 
\item Basic trading strategies show that the volatility predictions can reduce risk. 
\end{highlights}


\end{frontmatter}



\section{Introduction}



\begin{figure}[h!]
    \centering
    \includegraphics[width=\textwidth]{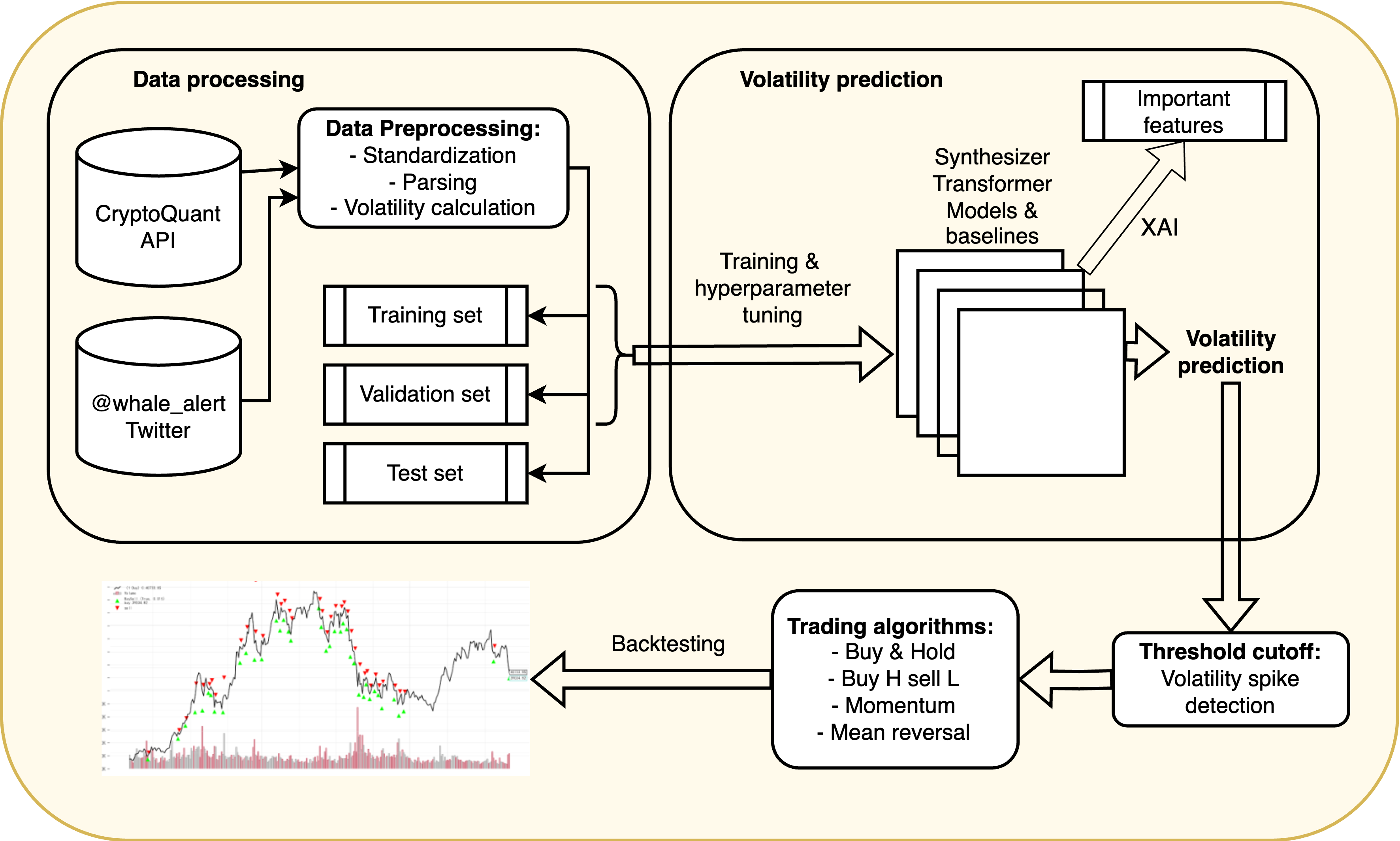}
    \caption{Overview of the proposed study. }
    \label{fig:overview}
\end{figure}

This paper studies the most popular cryptocurrency, Bitcoin, which is currently traded on more than 500 exchanges. Since Bitcoin is the first cryptocurrency, established in 2008~\citep{nakamoto2008Bitcoin}, it provides the longest historical data to study. Compared to traditional financial instruments like equities and commodities, cryptocurrencies like Bitcoin have large, so-called `whale' holders, which consist of about 1,000 people who own around 40\% of the market~\citep{kharif2017Bitcoin}. In this paper, we explore how large Bitcoin transactions from these whales affect the market volatility. We propose a state-of-the-art deep learning Synthesizer Transformer model \citep{tay2020Synthesizer} that predicts if Bitcoin's volatility will be extreme the next day, based on transaction data from these whales as well as a variety of features from CryptoQuant, including on-chain metrics, miner flows, and more. We compare this proposed model with existing baseline models and propose a simple trading strategy to demonstrate the practical usefulness of the predictions. In our experiments, we also analyse the importance of the different CryptoQuant and whale-alert features that most influence volatility. An overview of our paper is provided in Figure~\ref{fig:overview}. The code of our proposed (trained) models is made available online\footnote{\url{https://github.com/dorienh/bitcoin_synthesizer}}.

We focus on the volatility of Bitcoin as this digital asset dominates the cryptocurrency market with the largest market cap after USDT. In finance, volatility refers to the degree of variation of an asset's price over time~\citep{black2012dictionary}. Market volatility is generally considered a vital metric to evaluate the level of risk, and thus it plays a critical role in assessing the stock market risk and the pricing of derivative securities~\citep{yang2020html}. Compared to traditional financial instruments, the price of Bitcoin is highly volatile~\citep{blau2017price}. 
In general, the Bitcoin market is currently highly speculative, and thus more susceptible to speculative bubbles than other traditional currency markets \citep{grinberg2012Bitcoin, cheah2015speculative}. 
Bitcoin has recently also found its place in portfolios to hedge against the global geopolitical crisis  \citep{dyhrberg2016bitcoin} and reduce financial market uncertainty (\citep{platanakis2019portfolio, fang2019does, colon2021effect}, hence studying risk and assessing exposure is important to cryptocurrency investors, and it becomes important to model and forecast the volatility of Bitcoin. In this paper, we focus on predicting future spikes in Bitcoin's volatility.


This study aims to gain further insights into the market conditions that may cause drastic increases in volatility in Bitcoin markets. Our contribution is threefold. We first thoroughly explore both CryptoQuant data and the influence of whale transactions on volatility. Second, we propose and evaluate a state-of-the-art Synthesizer Transformer model to predict volatility. Finally, we propose a basic trading strategy that leverages the volatility predictions to reduce downward risk. We briefly touch upon the importance of these contributions in what follows.

First, in this study, we gather a dataset from CryptoQuant\footnote{\url{http://cryptoquant.com}}, as well as whale transaction tweets from January 2018 to September 2021. The former includes information such as exchange and miners transactions as well as liquidations and open interest caused by trading with leverage (full feature set, see Table~\ref{tab:cryptoquant_table}). We thoroughly explore the relationship between this data and Bitcoin's next-day volatility, and focus on discovering large market movements induced by the ripple effects of large whale transactions and on-chain movements.  


Second, we propose a Synthesizer Transformer model to perform the volatility spike prediction. The Transformer architecture has proven to be extremely efficient for a range of tasks related to time series such as text translation \citep{vaswani2017attention}, music generation \citep{makris2021generating}, emotion prediction from movies \citep{thao2021attendaffectnet}, and speech synthesis \citep{li2019neural}. In finance, it has been shown to be efficient at stock price \citep{liu2019Transformer, zhang2022Transformer} and even stock \textit{volatility} prediction \citep{yang2020html}. In the cryptocurrency markets, we see that it has been used for Dogecoin \citep{sridhar2021multi} and Bitcoin \citep{jain2019cryptocurrency} price prediction. In this work, we expand the existing literature by including CryptoQuant and whale data (plus technical indicators calculated on this data). We then go beyond just building a black-box model, but also explore the influence of these features on volatility prediction through explainable artificial intelligence (XAI) techniques with the Captum library \citep{kokhlikyan2020captum}. Instead of using Vanilla (standard) Transformer architectures, we change the typical dot product self-attention mechanism to Synthesizer attention, which learns synthetic attention weights without token-to-token interactions. By doing so, we optimize the attention span of the model. Recent work has shown that Synthesizer Transformers outperform traditional Transformers. Even a simple Random Synthesizer has shown to be 60\% faster than a traditional Transformer \citep{tay2020Synthesizer}.
In an experiment, we compare our proposed architecture to other configurations and baseline traditional models like GARCH. We show that it is a useful and reliable method for forecasting volatility in cryptocurrencies.

Finally, we explore the usefulness of our predictions by backtesting a number of trading strategies that use the predicted volatility. 
In practice, investors often use volatility to trade derivative instruments such as put and call options \citep{ni2008volatility}. Since it is hard to backtest such a strategy in a Bitcoin context, we propose examples of simple trading strategies which use trading signals based on our volatility prediction model. 
We explore four different strategies: buy \& hold, buy-low-sell-high, mean reversion and momentum-based. When we include position scaling based on volatility, we notice an increase in the cumulative returns as well as the Sharpe ratio. In future work, these strategies should further be improved, but for now, they serve as a simple example that our prediction model can be used to lower the downside risk of a portfolio. 

The rest of this paper is structured as follows. In Section~\ref{sec:2}, we review the existing literature, followed by a thorough description and visualisation of the dataset that was collected. Next, the proposed Synthesizer Transformer models are introduced in Section~\ref{sec:4}. Section~\ref{sec:5} provides a detailed account of the performance of the volatility prediction models compared to benchmarks, as well as insight into the important features through XAI. The setup and results of the backtesting experiment is described in Section~\ref{sec:6}. Finally, we provide conclusions and suggestions for further work in Section~\ref{sec:7}.

\section{Literature Review}
\label{sec:2}
We provide a brief overview of literature related to on-chain data, using Twitter data for volatility and price prediction, followed by deep models for cryptocurrency-related predictions. For a more complete overview, the reader is referred to \citep{zou2022multimodal, charandabi2021prediction, khedr2021cryptocurrency, charandabi2022survey}.

\subsection{Cryptocurrency-specific data}


The cryptocurrency markets are fundamentally quite different from traditional stock markets. One of the key differences is the transparency provided by blockchain technologies \citep{biswas2019analysis}. Transparency is one of the key features of Bitcoin trading as the entire trading history is available and traders are provided with information on the complete state of the order book, but trading itself is pseudonymous. This transparency provides unique features that may be useful for price and volatility prediction. 

On-chain data includes information from the blockchain ledger, such as the details of each transaction (e.g. from which wallet, to which wallet, amount, fees paid to miners), and the difficulty of mining blocks as well as the block sizes \citep{jagannath2021chain, kim2022deep}. The availability of such data can gives us incredible insight in upcoming price movements \citep{zheng2021chain}. The transparency in the blockchain even allows us to access the entire transaction history ever recorded. There is no hidden volume (as in iceberg orders) nor dark pools~\citep{dimpfl2017bitcoin}. However, to use this data would require a huge amount of computing power, hence, we focus on aggregated on-chain data instead. CryptoQuant provides us with a wide selection of such features, and also includes exchange data such as the amount of liquidations, as well as data on Bitcoin miners. 


Looking at existing literature, we see that utilizing this transparency allows one to establish a trader's edge. For instance, \citet{kim2022deep} show that on-chain data can be useful when predicting Bitcoin's price with a self-attention-based multiple long short-term memory model (SA-LSTM). While they provide a list of 42 variables used, there is no ablation study or XAI method used to identify which variables are most important. \citet{jagannath2021chain} equally show that the Ethereum price can be predicted using on-chain data and a self-adaptive LSTM model. A correlation analysis using their data reveals important correlated on-chain features to the price of Ethereum. These features include transaction rate, supply in smart contracts, block difficulty and hash rate. 

On-Chain data is not only useful for \textit{price} prediction, the correlation between on-chain transaction activities and \textit{volatility} has been shown by \citet{gkillas2021transaction}. \citet{raheman2021architecture}'s developed agent for crypto-portfolio management also uses on-chain data for price trend and volatility prediction. The literature available on the effects of various cryptocurrency-specific data such as on-chain data is still in its early shoes. In this work, we aim to not just build a predictive model for volatility, but also thoroughly analyse the patterns within the data and provide an XAI interpretation of the resulting model. 

In addition to CryptoQuant data, we also parsed a new dataset of whale transactions. An overview of the literature related to this is provided in the next subsection.


\subsection{Importance of Twitter data for volatility}

The CryptoQuant data offers us nice insights into aggregated on-chain data, miner data and more. It does not, however, include transactions by so-called `crypto-whales', holders of very large wallets. It is well known that cryptocurrencies are very volatile in nature, thus creating both outstanding benefits as well as a huge risk to investors~\citep{bariviera2017some, klein2018Bitcoin}. Part of this volatility can be attributed to large (whale) transactions and their ripple effect on the market. 
In this work, we will be using very specific Twitter content, namely `whale-alert' tweets. The Twitter account\verb|@whale_alert|, is a third-party information provider that ``monitors millions of daily cryptocurrency transactions and publishes notable events on Twitter in near real-time'' \citep{saggu2022intraday}. \citet{scaillet2020high} found a correlation between their `whale index' and high-frequency price jumps of Bitcoin.

Social media sources such as Twitter have been shown to be helpful data sources for stock or cryptocurrency price predictions. To name a few examples, \citet{lamon2017cryptocurrency} study whether including sentiment analysis of news and social media can improve models when predicting the price of Bitcoin and Ethereum. \citet{aharon2022Twitter} explore the relationship between two novel Twitter-based measures of economic and market uncertainty and the performance of four major cryptocurrencies. 
\citet{zou2022multimodal} shows that using BERT context embeddings of tweets with an LSTM model can improve Bitcoin price prediction. News and social media data have also been shown to be useful for \textit{volatility} prediction, as \citet{sapkota2022news} predicts Bitcoin volatility based on news sentiment, and \citet{akbiyik2021ask} use temporal convolutional neural networks for Bitcoin volatility prediction with Twitter sentiment. \citet{shen2019does} show that the number of tweets is a major determinant of the next day’s trading volume and realised volatility of Bitcoin. Finally, \citet{wu2021does} reported that there is a significant Granger-causality from Twitter-based uncertainty measures to Bitcoin, Ethereum, Litecoin, and Ripple prices in different time periods. In this work, we will focus on integrating tweets by \verb|@whale_alert| into our Transformer model.

\subsection{Deep neural networks for financial time series predictions}

Traditional models, like Generalised autoregressive conditional heteroscedasticity (GARCH)-based models) are widely used for volatility forecasting~\citep{engle1982autoregressive,bollerslev1986generalized}.  \citet{katsiampa2017volatility} and \citet{bergsli2022forecasting} study volatility forecasting for Bitcoin using GARCH and its variants. \citet{naimy2018modelling} concluded, however, that the predictive ability of GARCH is not good in the context of unusually high volatility, and performs better when volatility is relatively low. \citet{vilasuso2002forecasting} brings up one of GARCH's major limitations where ``its memory is sometimes not long enough to capture the persistence of some shocks that are observed to last for a very long time''. 
\citet{jiang2022forecasting} propose a time-varying mixture model, which includes an accelerating generalized autoregressive score (aGAS) technique into the Gaussian-Cauchy mixture (TVM)-aGAS model for forecasting Value-at-Risk for cryptocurrencies. 
Recently, however, many researchers have turned to ever more powerful deep learning models for financial time series prediction.

Just like in the stock market \citep{ding2015deep, hu2021survey, jiang2021applications}, deep learning models have become popular tools for price prediction in cryptocurrency markets \citep{zou2022multimodal, yao2018predictive, patel2020deep, akyildirim2021prediction, alessandretti2018anticipating, khedr2021cryptocurrency}. Looking at time series in general, recurrent neural networks, such as long-short term memory models (LSTMs)~\citep{hochreiter1997long} and gated recurrent unit (GRUs)~\citep{chung2014empirical} have been widely used for forecasting. 
When it comes to volatility prediction, \citet{vidal2020gold} proposed an architecture based on convolutional neural networks (CNNs) and long-short term memory (LSTM) units to forecast gold volatility. LSTMs were also used by \citet{jung2021forecasting} to forecast currency exchange rate volatility. Finally, temporal convolutional neural networks have been used with Twitter sentiment data to predict Bitcoin volatility \citep{akbiyik2021ask}.

In recent years, with the invention of the Transformer network \citep{vaswani2017attention}, deep models for time series prediction have become even more powerful. Transformers use a self-attention mechanism, to give relative focus on the context of an element of a time series, and are better able to capture long-term trends. In finance, we have seen the successful use of Transformer architectures for tasks such as stock price prediction \citep{ding2020hierarchical}, stock volatility prediction \citep{ramos2021multi}, and even cryptocurrency price prediction such as Dogecoin \citep{sridhar2021multi} and Bitcoin \citep{jain2019cryptocurrency}. The work on volatility prediction for Bitcoin with Transformers is relatively non-existent, except for the work by \citet{sapkota2022news} who built a model based on Twitter sentiment data. In this work, we explore how we can use the powerful Transformer architecture to perform Bitcoin volatility prediction, not only based on candlestick data, but also CryptoQuant data and whale-alert tweets. In addition, we implement the Synthesizer Transformer, to further optimize the attention mechanism.


\section{Dataset collection and analysis}
\label{sec:3}

The Bitcoin market provides interesting conditions from a volatility point of view. There is 24-hour continuous trading, 365 days a year, with a lack of central authorities (e.g., central banks), resulting in the absence of a volatility trading halt, and no pre-market/post-market  trading as compared to the equities market~\citep{brandvold2015price}. These market conditions, along with the complete transparency of the on-chain trading data, create an interesting opportunity for us to study the influence of different factors on volatility. To do so, we have gathered a dataset from January 2016 until September 2021, which consists of CryptoQuant (on-chain data and market data from cryptocurrency exchanges), and whale transaction tweets.  We will start below by discussing the features in this dataset and how we gathered them, and then move on to include technical indicators and data preprocessing.

\subsection{Data sources}

In this section, we discuss how we gathered whale transaction data which includes many aspects such as whale accumulation, whale dumping, miners' inflow and outflow, as well as exchanges' inflow and outflow.

\subsubsection{Whale-alert data}

Crypto `whales' include some of the largest wallet holders, and hence have a significant influence on both price and volatility \citep{nguyen2018factors}. In any volatility model, it is thus essential to include data about whale transactions. In order to do so, we tracked the Twitter handle \verb|@whale_alert|, which provides continuous alerts as whale transactions happens. Some example tweets by this handle are shown below: 

\begin{itemize} \setlength\itemsep{0em} \setlength{\parskip}{0pt}%
    \item ``\verb|997 #BTC (6,269,280 USD) transferred from #Bitfinex| 
    
    \verb|to Unknown wallet|''
    \item ``\verb|11,000 #ETH (2,473,411 USD) transferred from Unknown|
    
    \verb|wallet to #Gemini|''
    
    \item ``\verb|6,000,000 #USDC (6,000,000 USD) burned at USDC Treasury|''
\end{itemize}

Once we collected all of the tweets from 12 September 2018 (earliest available) to 18 October 2021, we filtered transactions using the hashtag \#BTC and the keyword `transferred', resulting in a total of 52,787 tweets. We then wrote a parser that uses a set of rules to obtain useful data from these tweets such as total daily inflow and outflow of wallet to exchange, e.g. the word after `from' will be the source of transaction and the word after `to' will be the destination. For all of the tweets gathered in a day, we determine the overall net transaction outflow or inflow of wallets to exchanges in one day, resulting in the following daily features: 
\begin{description} \setlength\itemsep{0em} \setlength{\parskip}{0pt}%
    \item [BTCminus] The amount of Bitcoin flowing out of wallets into exchanges.
\item [BTCplus] The amount of Bitcoin flowing into wallets from exchanges.
\item [USDminus] The amount of USD flowing out of wallets into exchanges.
\item [USDplus] The amount of USD flowing into wallets from exchanges.
\end{description}

This data is relevant for our task: a transaction from wallet to exchange typically indicates a bearish sentiment given that the seller is closing their Bitcoin position and may want to exchange it into fiat currency. On the other hand, a transaction from exchange to wallet means that a buyer is planning to keep their Bitcoin position (or at least not exchange it into fiat) and is therefore bullish. For the purpose of this study, we only examine Bitcoin transactions that flow either from exchange to wallet or wallet to exchange. The total net flow of transactions from wallet to wallet and exchange to exchange is ignored.

\begin{figure}[h!]
  \centering
    \includegraphics[width=0.8\textwidth]{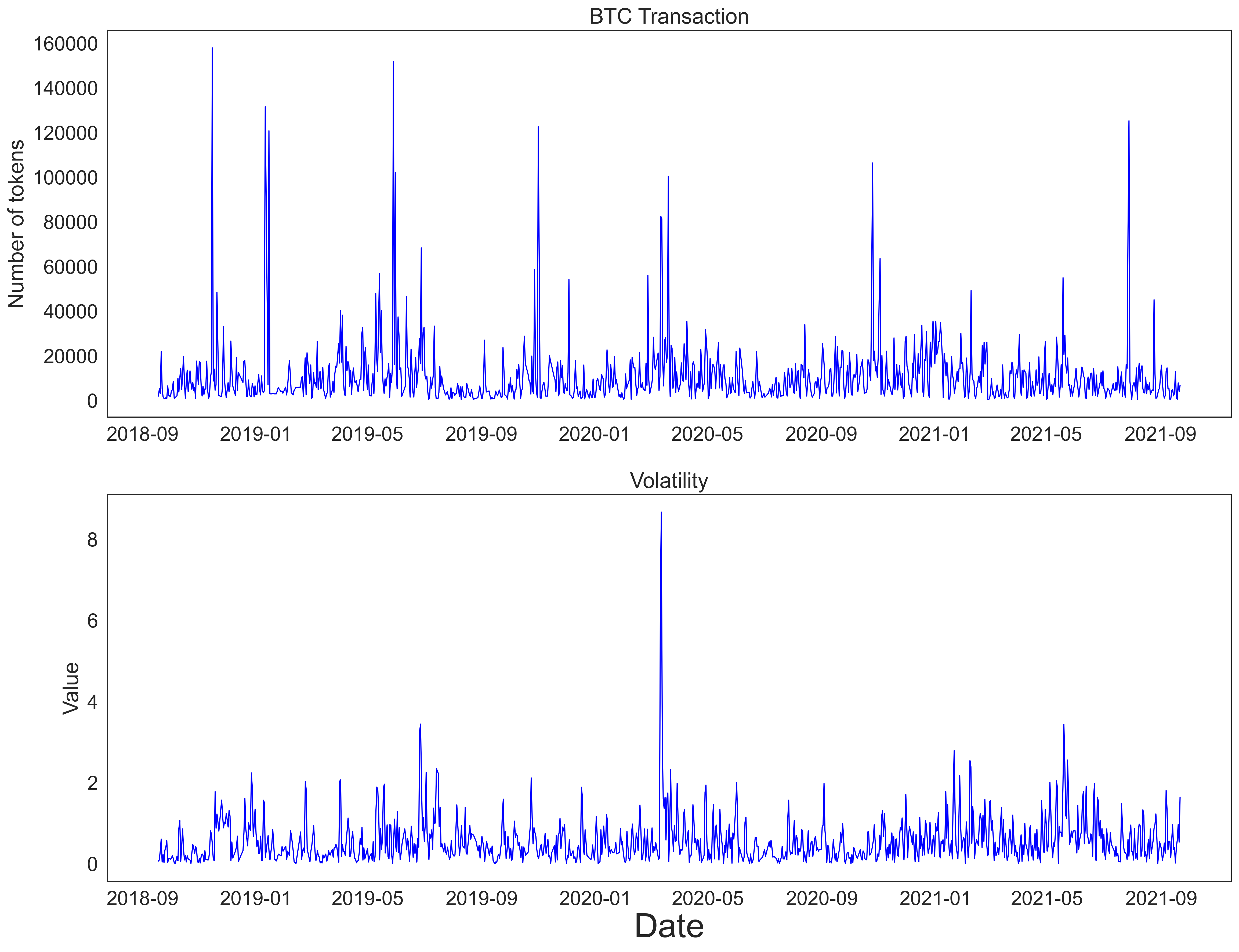}
    \caption{The daily amount of net BTC that flowed to or from exchanges per day, calculated as $abs(\text{BTCplus} - \text{BTCminus})$ (top). The BTC price volatility (bottom). }
    \label{fig:whale_alert}
\end{figure}

Figure~\ref{fig:whale_alert} plots the BTC price volatility against the number of BTC transactions measured in the daily amount of BTC that flowed to and from exchanges as per our whale-alert tweets. We see that there are patterns where volatility spikes during a spike in BTC transactions. There are 330 volatility spikes in total and we see that the net daily amount of BTC that flowed to or from exchanges (calculated as $abs(\text{BTCplus} - \text{BTCminus})$) has a Pearson correlation of 0.47 with daily BTC price volatility.


\subsubsection{CryptoQuant on-chain and exchange data}


CryptoQuant data provides comprehensive on-chain and market data gathered from both the blockchain as well as major cryptocurrency exchanges. Every single transaction that occurs in these markets is tracked by CryptoQuant. CryptoQuant even keeps track of which addresses are exchanges or mining pools, and aggregates the amount of BTC flowing between different types of entities, such as miners, and exchanges. In this study, we use CryptoQuant's\footnote{\url{https://cryptoquant.com/docs}} API to gather BTC related data. 
While a full overview of all the features we use is provided in Table~\ref{tab:cryptoquant_table} based on CryptoQuant's documentation\footnote{\url{https://dataguide.cryptoquant.com/}}, we elaborate on a few specific examples below:

\begin{description} 
    \item[miner\_inflow\_mean\_ma7] The 7-day moving average of miner inflow gives us insight into when whale accumulation occurs. Miners are often considered to be the original whales, as they typically hold large wallets. 

    \item[mtoe\_flow\_total] The miners-to-exchanges feature will keep track of how much BTC miners are transferring to exchanges. Typically, the main reason to send Bitcoin to an exchange would be to sell it, hence this can be a bearish indicator. 

    \item[miner\_outflow\_top10] The amount of Bitcoin that flows out of the 10 largest Bitcoin wallets held by miners. These whale wallets will be responsible for downward pressure and increased volatility if this variable increases. 




    \item[long\_liquidation] The amount of leveraged positions in BTC that were forced to exit due to volatility. High values for this variable hence often go hand in hand with high volatility. 

\end{description}


\subsection{Technical Indicators}

In order to improve the prediction of our volatility prediction model, we include some traditional technical indicators as input which have shown to be correlated to volatility~\citep{liashenko2020fractionally}. These include Exponential Moving Average (EMA), High-Low Spread, and Close-Open Spread. Exponential moving average indicators place a higher weighting on recent data compared to old data, hence, they are more reactive to the latest price changes compared to simple moving averages (SMAs). For this reason, we chose to include the 10th day EMA of the closing price instead of its SMA. This was calculated as per the below Equation~\ref{eq:ema} whereby $n$ is the number of days over which the EMA at time $t$ for a time series $X$ is calculated. The variable $S$ represents a smoothing factor, which we set to 2 for our study. 

\begin{equation}
  \text{EMA}_t = X_{t} \times (\frac{S}{1 + n}) + \text{EMA}_{t-1} \times (1 - \frac{S}{1 + n})
  \label{eq:ema}
\end{equation}

A second technical indicator is the High-Low Spread. This indicator gives insight into the intra-day total price movement. A higher value means that the price fluctuated in either direction in one day, thus indicating a higher volatility for that day, and vice versa.

\begin{equation}
  \text{High-Low Spread} = \frac{\text{High} - \text{Low}}{\text{Close}}  
\end{equation}

Finally, the Open-Close indicator provides a sense of the direction and size of the move. If the price goes up, this indicator will be negative, and vice versa. 

\begin{equation}
\text{Close-Open Spread} = \frac{\text{Close} - \text{Open}}{\text{Open}}
\end{equation}

\subsection{Data preprocessing}

\subsubsection{Missing values}

Some of the used technical indicators, such as exponential moving average, have a short warm-up period resulting in missing values. We can fill up the missing values by using the first available value since this only occurs at the very beginning of our (training) dataset.

The whale exchange tweets and derivatives data were only available from 2018 onwards. Before that period, we consider them to be zero. For leverage and derivatives data, it is easy to assume that the missing values are 0 since these assets were not yet available or created.

\subsubsection{Standardization}
Some of the distributions of the input features are skewed which would affect the Transformer's predictive abilities, hence we set out to standardize this. 
The descriptive statistics of features in Table~\ref{tab:before_normalization} are standardized in Table~\ref{tab:after_normalization}. Depending on how the data was skewed, we used five different techniques to standardize them as much as possible, as summarised in the latter table. 
We perform no change to features that are left skewed or that have a skewness less than 0.5 close to 0. As a default, for features with a higher right skewness (>0.5), we will perform a log() transform. In some cases, this can result in negative values, more specifically when the original values are <1, hence we cannot simply apply log(). We discuss the cases in which that happened and how we accounted for this: 

\begin{itemize} \setlength\itemsep{0em} \setlength{\parskip}{0pt}%
    \item 
MVRV, miner\_inflow\_mean\_ma7 and exchange\_mean\_ma7 have a skewness of 0.875, 1.55, and 1.16. Since all three of them have values in the range of 0.6 to 4, taking the logarithm would introduce negative values, therefore, we took the square root. 

\item 
The features HL\_sprd, miner\_inflow\_mean, exchange\_inflow\_mean, exchange\_outflow\_mean 
and miner\_outflow\_mean, have a higher maximum value (>5) and skewness (>3). Hence, we perform a slightly stronger transformation and take the cube root, so as the make the maximum values closer to 1.

\item 
Both the etom\_flow\_mean and mtoe\_flow\_mean features have a high skewness value of 17.19 and 20.48.
Since their minimum value is below 1 (0.0769 and 0.298), we first add a value of 1 to them and then take the logarithm. 

\item 
For the feature vol\_future, we took the power of $\frac{1}{4}$, to make the maximum value of 8.67 as close to 1 (threshold value) as possible. Given that this is our forecast variable, it was important to standardize this as good as possible. 

\end{itemize}



\subsection{Volatility}

\subsubsection{Calculating volatility}
We calculated the daily volatility $V_{annualised}$ for our dataset using the formula below. 

\begin{equation}
    log-returns = x_i = ln(\frac{C_i}{C_{i-1}})
\end{equation}
\begin{equation}
    V_{annualised} =   \sqrt{ \frac{\sum\limits_{i=1}^N (x_i-\mu)^2 \times 365 }{N}}   
\end{equation}

\textbf{where:}\\ 
\[\textit {$\mu$} =\text{mean of log-returns}\]
\[\textit {$C_i$} = \text{closing price of day $i$}\]
\[\textit {$C_{i-1}$}  = \text{closing price of day $i-1$}\]
\[\textit {$N$} =\text{number of days}\]

\begin{figure}[htbp!]
  \centering
    \includegraphics[width=.8\textwidth]{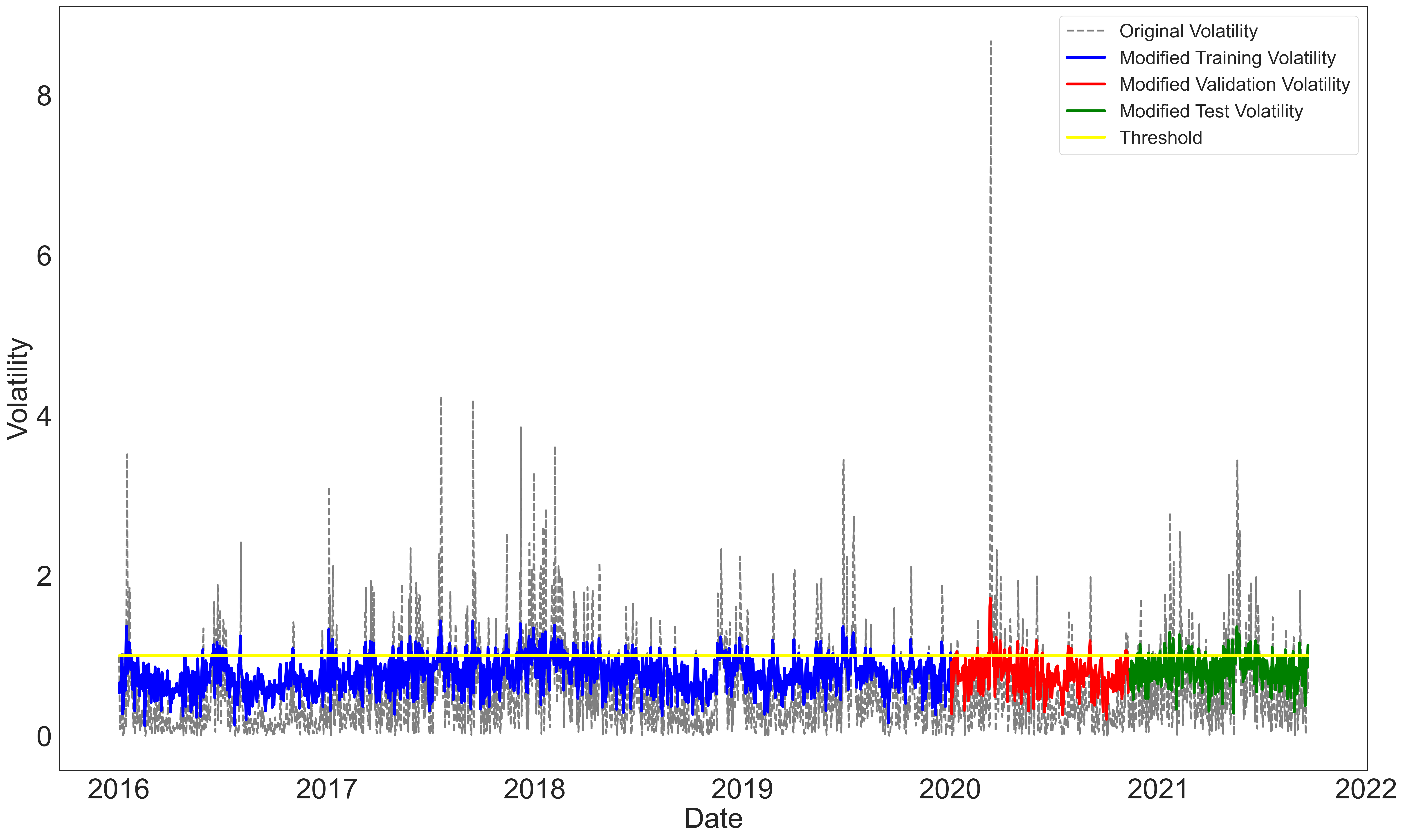}
    \caption{Daily Volatility before and after transformation.}
    \label{fig:volatility_transformation}
\end{figure}

\begin{figure}[htbp!]
  \centering
    \includegraphics[width=.8\textwidth]{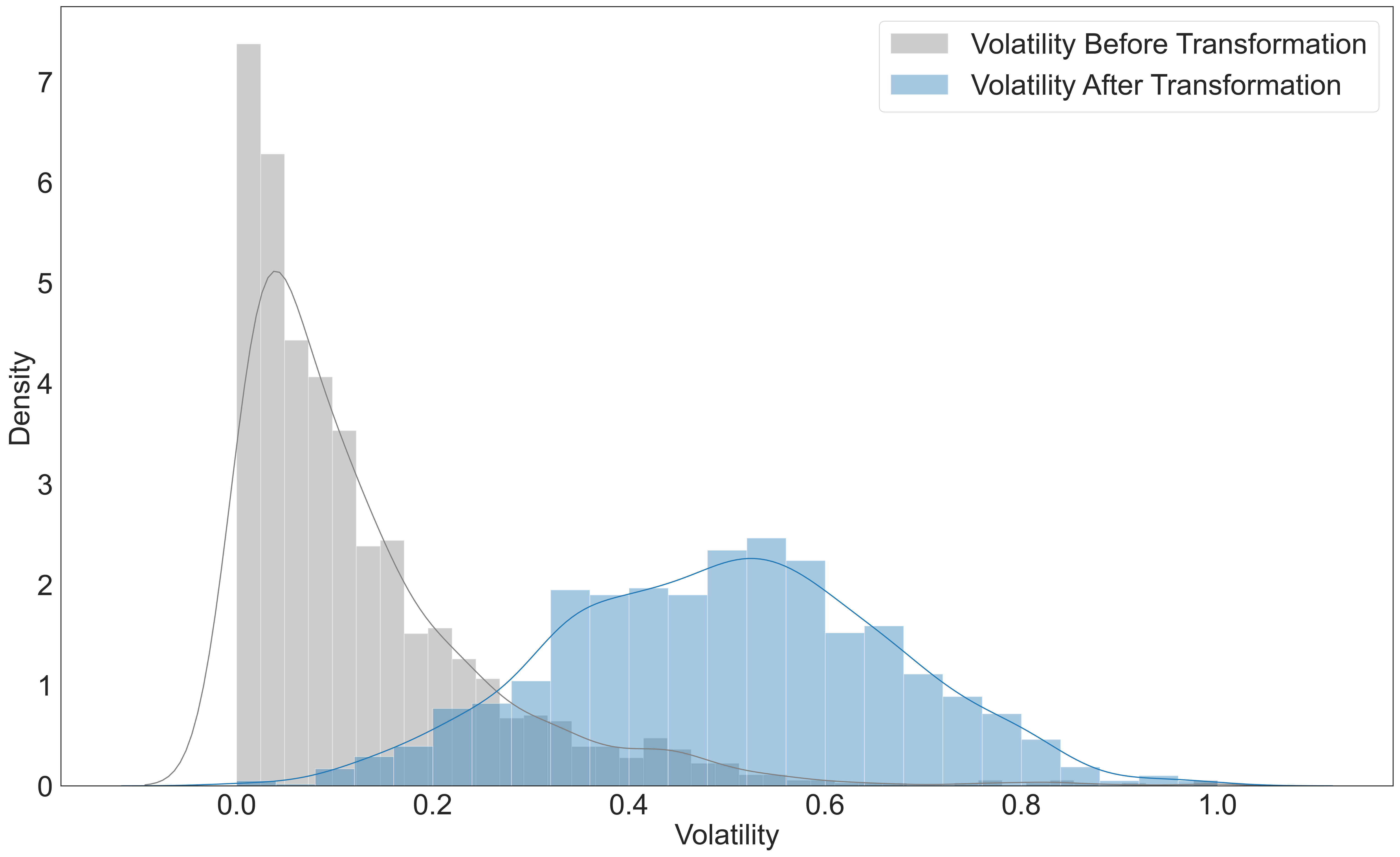}
    \caption{Volatility distribution before and after transformation.}
    \label{fig:volatility_distribution}
\end{figure}

As shown in Figure~\ref{fig:volatility_transformation}, the daily volatility in our dataset is in the range of 0.000234 to 8.67. This results in a long-tailed distribution (see Figure~\ref{fig:volatility_distribution}) with a skewness of 3.35. Since statistical learning models typically work better with normally distributed data, we apply a transformation to the volatility data by taking the power of $\frac{1}{4}$. This results in a distribution with a skewness of -0.001 and a volatility range of volatility of 0.124 to 1.72.

\subsubsection{Volatility spikes}

As shown in Figure~\ref{fig:volatility_transformation}, we classify days with a volatility $\ge$ 1.0 and with positive log-returns of the closing price as a \textit{volatility spike}. We set this volatility threshold to 1, because after applying the preprocessing transformation to the volatility (taking the power of $\frac{1}{4}$), all of the high volatilities with a magnitude > 1 will still be greater than 1 even though their magnitude has shrunk, and all of the low volatilities with a magnitude < 1 will still remain < 1.  There are 232 volatility spikes in the training set and 38 volatility spikes in the validation set. In the test set, there are 60 volatility spikes.

\subsubsection{Feature correlation with volatility}

To explore which of the (input) features from our dataset may be most correlated with the next-day volatility, and thus most important for our predictive model, we calculated several correlation metrics. Table~\ref{tab:corr} shows the R$^2$, and the Pearson as well as the Spearman correlation coefficients. We can see from the table that some features, such as volume, exchange\_inflow\_total and High\_Low\_Spread show a high correlation with the volatility. This indicates that these features will likely be important to improve our model's predictive power. This will later be verified by doing a Captum analysis in Section~\ref{tab:captum_feature} to explore the importance of each feature in our predictive model.





\section{Proposed Synthesizer Transformer}
\label{sec:4}

In this paper, we leverage a new type of Transformer, the Synthesizer Transformer~\citep{tay2021Synthesizer}. To properly understand our architecture, we first provide an overview of the Vanilla Transformer architecture upon which our proposed model is based.

\subsection{Transformer architecture}
\label{sec:model}

\begin{figure}[htbp!]
    \centering 
      \begin{subfigure}[t]{.49\linewidth}
    \includegraphics[width=0.8\textwidth]{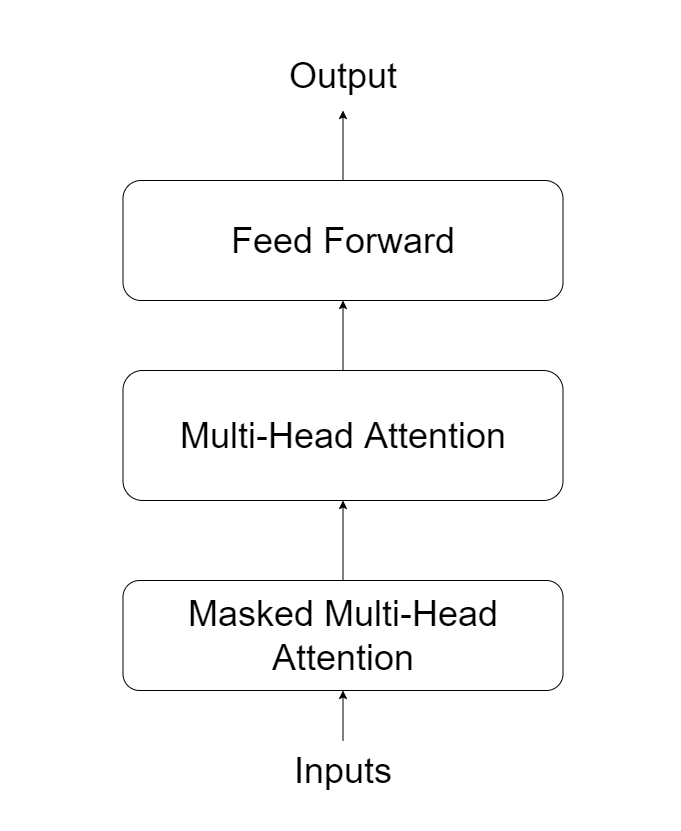}
    \caption{The proposed decoder-only Transformer architecture inspired by \citep{radford2018improving}.}
    \label{fig:Transformer_architecture}
      \end{subfigure}
      \begin{subfigure}[t]{.49\linewidth}
    \includegraphics[width=1\textwidth]{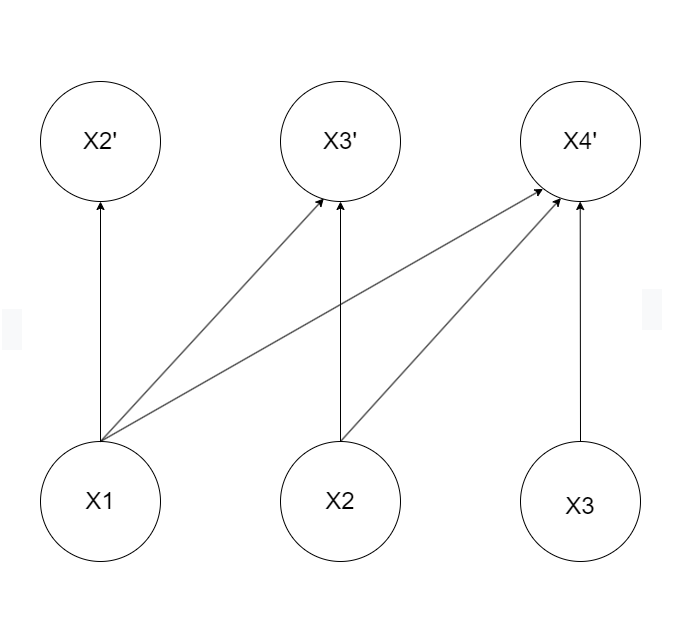}
    \caption{Transformer self-attention flows. The arrows indicate which inputs are received for making each prediction based on a time series $X$. }
    \label{fig:Transformer_attention}
      \end{subfigure}
      \caption{Insights into the used Transformer architecture.}
\end{figure}

The architecture used in this paper draws inspiration from the Generative Pre-trained Transformer 2 (GPT-2)'s decoder-only Transformer \citep{radford2018improving}, as shown in the Figure~\ref{fig:Transformer_architecture}. In this architecture, the input to the Transformer is a multivariate time series. The decoder takes the masked target sequence so that at each time step the decoder can attend to the previous $i$ time steps. This is illustrated in Figure \ref{fig:Transformer_attention} where the first input $X_1$ will result in a prediction for the next time step: $X_2$'. In the next step, the decoder is given the ground truth $X_1$ and $X_2$ values to predict $X_3$' and so forth. Therefore, at every new step, the model receives all the true inputs prior to predicting its next output, whereby each output token contributes equally to the training loss.

\begin{figure}[htbp!]
    \centering
    \includegraphics[width=1\textwidth]{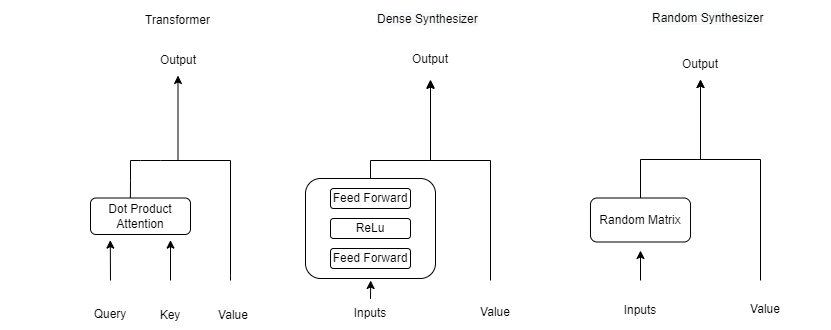}
    \caption{Types of self-attention mechanisms. On the left side, the traditional self-attention mechanism is depicted. The Dense and Random Synthesizer attention mechanism are shown next to it. Figure inspired by \citet{tay2021Synthesizer}.}
    \label{fig:self_attention_mechanism}
\end{figure}

For every output token, the self-attention score measures the importance of looking at each of the tokens previously seen in the sequence, for predicting the current token. In this traditional attention model (left in Figure \ref{fig:self_attention_mechanism}), the formula to calculate the attention score is provided in Equation~\ref{eq:attention}, and involves computing the dot product between the query vector ($Q$) and the key vector ($K)$ of the current token. For details, the reader is referred to \citet{radford2018improving}.

\begin{equation}
    \textit{Attention(Q, K, V)} = softmax(\frac{QK^T}{\sqrt{d_k}})V
    \label{eq:attention}
\end{equation}

whereby $\sqrt{d_k}$ represents both the dimension of the key vector $K$ as well as the query vector $Q$, and $V$ is the value vector.





\subsection{Synthesizer Transformer}

 \citet{tay2021Synthesizer}'s Synthesizer Transformer  is able to learn attention weights synthetically, without token-token interaction. This increases the speed of the Transformer by up to 60\%. Synthesizer Transformers can do this by removing the notion of query-key-values in the self-attention calculation and instead directly synthesizing the attention matrix. This is done using input $X_{h,l} \in\mathbb{R}^{N \times d}$ , where $h$ is the number of heads, $l$ is the sequence length and $d$ is the dimensionality of the model. This eliminates the need to calculate the dot product attention as described in the previous subsection. In their original paper,  \citet{tay2021Synthesizer} propose several synthetic attention variants, in this work, we implemented some of the best performing variants: dense, random, both of their factorized version, as well as a combination of dense and random with the Vanilla Transformer attention.

\noindent\textbf{Dense}

This type of dense synthetic attention uses a two-layer feed-forward network with ReLU activation to replace the traditional dot product attention. The attention matrix is simply learned by the dense neural network.

\begin{equation}
     \textit{Attention(V)} = Softmax(W_{2}(\sigma_{R}(W_{1}(X_{h,l})))V
\end{equation}

whereby $W_{2}$ and $W_{1}$ are feed-forward layers and $\sigma_{R}$ is a ReLu function
\newline

\noindent\textbf{Random}\\

The random synthetic attention mechanism does not rely on pairwise token interactions or any information from individual tokens. This way, it aims to capture a global task-specific alignment that obtains good results across a large number of samples. The attention is calculated as follows: 

\begin{equation}
     \textit{Attention(V)} = Softmax(R)V
\end{equation}

whereby $R$ is a randomly initialised $N \times N$ matrix. The weights in this matrix are then optimized during training. 
\newline

\noindent\textbf{Factorized Dense and Random}

The number of parameters added to the network in the above variations is $d \times N$ and  $N \times N$ respectively. When the sequence length is large, these synthetic attention models can be slightly harder to train. Hence, we also included factorized variations, which allow the models to perform competitively in practice. In addition, this form of attention also seems to help prevent overfitting. For details on how to calculate attention the reader is referred to \citet{tay2021Synthesizer}.






\noindent\textbf{Mixture dense and random}

All of the proposed synthetic attention variants can be mixed in an additive fashion. This results in mixture Synthesizer Transformers (mix). In this work, we experiment by mixing a dense Synthesizer Transformer and a Vanilla Transformer (mix dense) as well as a random Synthesizer and Vanilla Transformer (mix random). The resulting attention is calculated as the sum of the attention calculated by the Vanilla Transformer and the selected Synthesizer Transformer's attention.






\section{Volatility prediction}
\label{sec:5}

\subsection{Experimental setup}




We conduct a thorough experiment to evaluate the performance of the volatility prediction Synthesizer Transformer models (with different attention mechanisms) and compare it to existing baseline models: Vanilla Transformer, LSTM, and GARCH. We first perform hyperparameter optimization using the validation set. The final results using the best parameters are reported on the test set. After finding the best model, we use Captum, a PyTorch library for model interpretability, to identify the input features that contribute most to the prediction result. 

We evaluate the models on two tasks: predicting calculated volatility (regression) and predicting volatility spikes (classification). The latter is accomplished by converting the predicted volatility values into two classes: `volatility spike' and `non-volatility spike'. A prediction is considered to be a volatility spike when the predicted volatility is greater than or equal to 1 and the log-returns were positive, otherwise we label it as `non-volatility spike'.


\subsubsection{Training-Test Split}

We train our models using the dataset described in Section~\ref{sec:3}, split into a training, validation, and test set as described below: 

\begin{itemize} \setlength\itemsep{0em} \setlength{\parskip}{0pt}%
\item Complete dataset: 02/01/2016 to 21/09/2021 (2,090 days).
\item Training set: 02/01/2016 to 02/01/2020 (1,462 days) (70\%).
\item Validation set: 03/01/2020 to 11/11/2020 (314 days) (15\%). 
\item Test set: 12/11/2020 to 21/09/2021 (314 days) (15\%).
\end{itemize}

There are a total of 232 volatility spikes in the training set and 38 volatility spikes in the validation set. In the test set, there are 46 volatility spikes. We should note that the non-stationarity of financial data is a known issue \citep{de2018advances}. Ideally, we would train and test with a rolling time frame over our entire dataset, however, due to the fact that the Transformer model needs as much data as possible, we use an out-of-time test set.



\subsubsection{Baseline comparison models}

Since we are working with a new dataset, there are no existing benchmarks available to directly compare our results to. In order to overcome this, we trained a few baseline models: a Vanilla Transformer, long-short term memory model (LSTM), and GARCH.

The Vanilla Transformer is the same architecture as our proposed Synthesizer Transformer, but uses the original attention mechanism as per Subsection~\ref{sec:model}. Secondly, long-short term memory models (LSTMs)~\citep{hochreiter1997long} are a type of recurrent neural network that are known for their ability to capture long-term dependencies in time series data as well as avoid the vanishing gradient issue~\citep{chuan2018modeling}. The full configuration of the networks used as baseline is described in Subsection~\ref{sec:tuning}. Finally, we also explore a statistical model often used in time series analysis: Generalized AutoRegressive Conditional Heteroskedasticity, or GARCH~\citep{li2002recent}. This model extends the Autoregressive Conditional Heteroskedastic Models (ARCH) model, by including a moving average component (ma) joint with the autoregressive component. This model is often used for volatility prediction, even for Bitcoin ~\citep{dyhrberg2016bitcoin}. As our baseline model, we use GARCH(1,1), which is the first order GARCH model using the ARCH library in Python \footnote{\url{https://github.com/bashtage/arch}}. 



\subsubsection{Evaluation Metrics}

We use several metrics to evaluate the volatility prediction models: root mean square error (RMSE), F1-score, precision, and recall. The first metric looks directly at the regression results, the others look at the resulting predicted volatility spikes (classification). 
For the regression evaluation, we opted to use RMSE as it is more sensitive to prediction errors with a large difference from the ground truth.




When evaluating volatility \textit{spike} prediction, we need to take into account that our (test) dataset is not balanced as there are fewer volatility spikes (60) than non-volatility spikes (254). We use precision to see how many correctly predicted spikes (TP) the model predicted correctly out of all predicted spikes (TP+FP). 

\begin{equation}
    Precision = \frac{TP}{TP+FP}
    \label{eq:precision}
\end{equation}

Recall complements precision by measuring how many spikes the model predicted correctly out of the actual spikes. 

\begin{equation}
    Recall = \frac{TP}{TP+FN}
    \label{eq:recall}
\end{equation}

In addition, the F1-score provides an integrated metric as the harmonic mean between precision and recall. Overall, a balance of high recall and high precision is preferred because it assumes that the model is well fitted, although it is possible to rely solely on either recall or precision depending on the use case. 

\begin{equation}
    F1-score = \frac{2*Precision*Recall}{Precision+Recall} = \frac{2*TP}{2*TP+FP+FN}
    \label{eq:f1}
\end{equation}


\subsection{Hyperparameter tuning and implementation details}
\label{sec:tuning}

We set the sequence length of all Transformer models to be 64 and the weight decay to be $1\text{e}^{-6}$. We train all the neural network models using Adam optimizer with an initial learning rate of $1\text{e}^{-5}$. All Transformer models use early stopping with the maximum number of epochs set to 10,000 and a patience of 200 to prevent overfitting. In addition, we use the validation set to finetune the models' hyperparameters as displayed in Table~\ref{tab:hyperparameters_tested}. The resulting best parameter settings with the lowest RMSE loss on the validation set are displayed in Table~\ref{tab:best_parameters}.

\begin{table}[h!]
\small
    \centering
    \begin{tabular}{lll}
    \toprule
        Feature & LSTM models & Transformer models\\
        \midrule
        Number of layers & 1, 2, 4, 8 & 1, 2, 4, 8\\
        Number of hidden layers & 16, 32, 64, 128 & NA\\
        Number of heads & NA & 2, 4, 8 \\
        Batch size & 4, 8, 16, 32, 64   & 4, 8, 16, 32, 64  \\
        Dropout & 0.1, 0.2  & 0.1, 0.2\\
           \bottomrule
    \end{tabular}
    \caption{An overview of the hyperparameters tested for different neural network architectures.}
    \label{tab:hyperparameters_tested}
    
\end{table}

\begin{table}[h!]
    \centering
\scriptsize
    \begin{tabular}{ll}
    \toprule
        Model & Best hyperparameter settings \\
        \midrule
        LSTM  & batch size=4, dropout=0.2, hidden layer=64, layers=8 \\
        Transformer (V) & batch size=4, dropout=0.2, heads=4, layers=2\\
        Synthesizer (R) &  batch size=4, dropout=0.2, heads=4, layers=4\\
        Synthesizer (FR) &  batch size=4, dropout=0.2, heads=4, layers=8\\
        Synthesizer (D) &  batch size=4, dropout=0.2, heads=8, layers=4\\
        Synthesizer (FD) & batch size=4, dropout=0.2, heads=4, layers=4\\
        Synthesizer (MD) &  batch size=4, dropout=0.1, heads=2, layers=4\\
        Synthesizer (MR) & batch size=4, dropout=0.1, heads=8, layers=2\\
           \bottomrule
    \end{tabular}
    \caption{The best hyperparameters based on the validation set, for the different Transformer models. We use R for random, F for factorised, M for mixed, D for dense, and V for Vanilla models.}
    \label{tab:best_parameters}
\end{table}

\subsection{Volatility prediction results}

The results for predicting next-day volatility are displayed in Table~\ref{tab:model_prediction_results}. The left column displays the RMSE for the regression problem (predicting next-day volatility). We then used a threshold $T$, to determine if a volatility spike was predicted. Our default value for $T$ is 1, and for this value we show the F1-score, precision, and recall in the table. We also included the number of True Positives and False Negatives for a few other thresholds to gain insight in how to improve the prediction certainty in Table~\ref{tab:threshold}. 

From the table, we can see that many of the proposed Synthesizer Transformer models perform well, both in terms of the F1-score (which is consistently above 0.377) as well as RMSE (which is close to 0.1). The baseline LSTM model as well as the Vanilla Transformer consistently perform worse with F1-scores of 0.1714 and 0.2857 respectively. We also ran a basic GARCH(1,1) model which does not perform very well. Since the predictions were too low, no spikes were detected, leaving the precision and recall as zero. We can speculate that GARCH is not the most appropriate model for our task definition. This is in line with the findings of \citet{naimy2018modelling}, who find that GARCH is not well suited in a high-volatility context. 

When comparing the different types of Synthesizer Transformers, the dense model has a slightly better performance, with the model with factorized dense attention obtaining 0.101 in RMSE and 0.4625 in F1-score. In general, the factorised models slightly outperform the non-factorized models in terms of the F1-score. 

\begin{table}[h!]
\hspace*{-0.8cm}
    \centering
    \scriptsize 
    \begin{tabular}{l|c|ccc|cccccc|cc}
    \toprule
            Model & RMSE & F1-score & Precision & Recall  &  TP & FN & TN & FP \\
            
            \midrule
            GARCH(1,1) & 0.303  &0.000 &0.000 &0.000 & 0 & 60 &254 & 0\\
            
            LSTM   &0.095 &0.171 &0.600  &0.100 &6 &54 & 250 &4\\
            
            Transformer (V) & 0.095  &0.286 &0.500 &0.200    &12 &48 &242 &12\\
            
            Synthesizer (R) &0.114   &0.374 &0.303 &0.500    &30 &30& 185& 69\\
            
            Synthesizer (FR)  &0.123   & 0.414 &0.316 &0.600   &36 &24& 176 &78\\
            
            Synthesizer (D)  & 0.103  &0.448 &0.405 &0.500  &30 &30 & 210& 44\\
            
            Synthesizer (FD)  &0.101   &0.463 &0.370 & 0.617   &37 &23& 191 &63\\
            
            Synthesizer (MD)  & 0.100  &0.385 &0.429 &0.350  &21 &39& 226& 28\\
            
            Synthesizer (MR) & 0.101  &0.400 &0.400 &0.400  &24 &36& 218& 36\\
           \bottomrule
    \end{tabular}
    \caption{Model Prediction Results for predicting volatility  as regression (RMSE), and as a classification task (F1-score etc.). The True/False Positive/negative (TFPN) results for predicting extreme volatility spikes are also displayed in the last columns. We use R for random, F for factorised, M for mixed, D for dense, and V for Vanilla models.}
    \label{tab:model_prediction_results}
    \hspace*{2cm}
\end{table}

We included different values for our classification threshold $T$ and reported TP and FN values in Table~\ref{tab:threshold}. We see that if we want to have a higher certainty for true positives and a lower chance of false negatives, then setting a higher threshold can help us achieve this. Looking at the Synthesizer (FD), a threshold of 1.2 can help us obtain a recall of 0.85714 (6/(1+6) compared to the original 0.370. This means that we correctly predict 6 our of 7 (larger) volatility spikes. Even with a threshold of 1.1, the Synthesizer Transformer correctly predicts more than 50\% of the volatility spikes.

\begin{table}[h!]
\hspace*{-0.8cm}
    \centering
    \scriptsize 
    \begin{tabular}{l|cccccc}
    \toprule
            Model   TP  & FN   & TP   & FN  & TP   & FN & TP \\
             &  \multicolumn{2}{c}{T $\ge$ 1.3}   & \multicolumn{2}{c}{T $\ge$ 1.2}  & \multicolumn{2}{c}{T $\ge$ 1.1}  \\
            
            \midrule
            GARCH &0 &1 &0 &7 &0 &29 \\
            
            LSTM   &1 &0 &2 &5 &4 &25 \\
            
            Transformer (V)    &1 &0 &4 &3 &6 &23 \\
            
            Synthesizer (R)   &1 &0 &5 &2 &18 &11 \\
            
            Synthesizer (FR)    &1 &0 &6 &1 &21 &8 \\
            
            Synthesizer (D)   &1 &0 &6 &1 &17 &12  \\
            
            Synthesizer (FD)    &1 &0 &6 &1 &21 &8  \\
            
            Synthesizer (MD)    &1 &0 &4 &3 &15 & 14 \\
            
            Synthesizer (MR)     &1 &0 &4 &3 &16 &13 \\
           \bottomrule
    \end{tabular}
    \caption{True positive (TP) and false negative (FN) when predicting extreme volatility spikes with different thresholds. We use R for random, F for factorised, M for mixed, D for dense, and V for Vanilla models.}
    \label{tab:threshold}
    \hspace*{2cm}
\end{table}













\subsection{Model explainability}

To gain insight into which features are important for predicting volatility, we used the Captum library for model interpretability. More specifically, we used the feature ablation function \citep{kokhlikyan2020captum} to understand important features that contribute to the prediction of each of the models. Table~\ref{tab:captum_feature} shows the top 3 features in terms of the absolute value of the weight attribute score based on the feature ablation attribution algorithm for each of the models. 

The absolute value of the score, informs us about the importance of this feature for predicting the next-day volatility. Some notable recurring features are important across different models based: taker\_buy\_volume, HL\_spread and volume. Looking back at the initial correlation analysis that we performed in Table~\ref{tab:corr}, we confirm the importance of HL\_spread and volume for volatility prediction since they have the highest correlation with vol\_future. 

The feature called taker\_buy\_volume refers to the volume of perpetual swap trades that market takers buy (and vice versa for taker\_sell\_volume). Being a `taker' indicates someone who buys or sells at the market price. When the takers' buy volume is much larger than the takers' sell volume, this indicates a bullish movement. Other important features include exchange\_outflow\_mean\_ma7 and exchange\_transactions\_count\_inflow. An increase in the latter indicates that more people are active in exchange flows which in turn indicates an increase in interest, leading to an increase in volatility.



Looking at the features that we extracted from Twitter, we find that our variables related to whale transactions also come out as being important with most of them listed as the 10th or 20th most important feature.  The most important is the USDminus, which is the 4th most important feature for the Synthesizer Transformer (FD) with an ablation score of -0.0398. This feature is also shown as the 12th most important feature for the Synthesizer Transformer (MD).




 \begin{table}[htbp!]

\hspace*{-1.6cm}
\scriptsize
\begin{tabular}{llclclc} 
 
 \toprule
Model & Feature 1 & Score & Feature 2 &Score& Feature 3 & Score\\ 

 \midrule

V &HL\_sprd&0.09&volume&0.07 &funding\_rates &-0.06\\

R &taker\_sell\_volume&0.09&HL\_sprd&0.09 &taker\_buy\_volume&0.07 \\

D &exchange\_outflow\_mean\_ma7&-0.07&close&0.05&HL\_sprd&0.05\\

FR &HL\_sprd &0.08 &taker\_buy\_volume & 0.08&taker\_sell\_volume &0.08 \\

FD &close&0.07&volume &0.05 &exchange\_transactions\_count\_inflow&0.05\\

MD &taker\_buy\_volume&0.08&taker\_sell\_volume &0.06 &volume &0.06  \\

MR &volume &0.08&HL\_sprd&0.06 &taker\_buy\_volume&0.06 \\

\bottomrule
\end{tabular}
\caption{The top 3 most important features of the Transformer models according to Captum's feature ablation function with their attribute score. \\}
\label{tab:captum_feature}
\end{table}


\section{Trading strategy experiment}
\label{sec:6}

In order to evaluate the usefulness of the volatility model, we implemented a few simple trading strategies that take signals from the volatility prediction model, and backtested them. It is worth noting that these strategies are very basic, and can undoubtedly be improved. They solely serve to show whether our predicted volatility metric can help increase our risk-adjusted profits.

\subsection{Backtesting strategies}

We used the predicted volatility (for each model) and used it as a signal for our strategies. For all of the strategies, we start with an initial capital of \$10,000. Each buy signal will be 5\% of the remaining capital, with pyramiding. Trading costs were set to 0.1\% for this experiment which is relatively higher than many exchanges. The backtesting was performed using the Backtrader library in Python\footnote{\url{www.backtrader.com}}. 


We test each of the strategies with and without \textbf{volatility scaling} for setting the position size. As explained above, the strategies typically open a position by buying a fixed percentage of total capital (5\%). With volatility scaling, they open a position by buying 5\% of capital $times$ volatility. This means that when the volatility is higher, we are trying to gain an edge by using a higher percentage of capital to open a position. \citet{hoyle2018volatility} suggest that volatility scaling can potentially improve the Sharpe ratio of the returns. The four strategies that we tested are described below. 

\subsubsection{Buy-and-hold}

This baseline strategy buys Bitcoin at the start and holds it until the very last day. Due to its constant market exposure, we can expect a higher risk, with, during long enough certain periods, higher returns.

\subsubsection{Buy-low-sell-high}

An often used strategy is to buy when prices are low, and sell when they are high. We modified this idea to buy when volatility is low ($V < 1$) and there is a decrease in log-returns, and sell when a volatility spike is detected ($V \ge 1$), regardless of the price.

\subsubsection{Momentum}

The proposed Momentum strategy will buy when a volatility spike is predicted and there is an increase in log-returns over the past 2 days. The position will close the next day.

\subsubsection{Mean Reversion}

The proposed Mean Reversion strategy will buy when a volatility spike is predicted and there is a decrease in the log-returns over the past 2 days. The position will close the next day.

\subsection{Evaluation metrics}

We used the following metrics to evaluate our backtesting experiment:

\begin{description} \setlength\itemsep{0em} \setlength{\parskip}{0pt}%
    \item[Time in market] - The number of days for which a position was open.




\item[Max. Drawdown] - The maximum observed loss from the maximum portfolio value to a subsequent through value before a new maximum is attained (in percentage).

\item[Kelly Criterion] - Determines the optimal theoretical positions size. 


\item[Daily VaR(\%)] - Daily Value-at-Risk. The VaR reflects the potential loss within a day and a certain confidence level (95\%).


\item[PnL] The total profit and loss in percentage.

\end{description}

\subsection{Backtesting results}

Table~\ref{tab:backtest_results} shows the result of our backtesting experiment. 
The Buy and hold strategy has a Profit and Loss (PnL) of 12.2\% for almost one year of holding. The disadvantage of such a strategy, is its constant market exposure, resulting in a high maximum drawdown of 13.66\%. Many investors may want to avoid such exposure and instead save fiat for bargain buying opportunities. The buy-low-sell-high strategy performs best in terms of PnL (24\%), especially with volatility scaling. This strategy, however, still has a very high time in the market, resulting in a max. drawdown ranging from  -15\% to -25\%. The Momentum strategy, on the other hand, shows a very low time in the market (less than  20\%), with a PnL between 2\% to 10\% for the different proposed Transformer models and a max. drawdown of less than 5\%. 


In general, profit increases when volatility scaling is used. The Sharpe ratio, Kelly criterion and PnL generally all increase when using volatility scaling for position size, compared to unscaled position sizing. The risk, however, also increases in terms of Daily VaR and max. drawdown. Hence, investors and traders have to weigh the cost and benefits of volatility scaling and see whether are they comfortable adding more risk to their strategy so as to profit more.


{\scriptsize \begin{longtable}{lcccccc}
\\ \toprule 
  & Time In & Sharpe &Max &Kelly &Daily & \\
Model  & Market(\%)&  Ratio& Drawdown(\%)& Criterion(\%)& VaR(\%)&PnL(\%) \\
 \midrule
\endfirsthead

\multicolumn{7}{c}%
{{-- continued from previous page}} \\ 
\toprule 
  & Time In & Sharpe &Max &Kelly &Daily & \\
Model  & Market(\%)&  Ratio& Drawdown(\%)& Criterion(\%)& VaR(\%)&PnL(\%) \\
 \midrule
\endhead

\bottomrule \multicolumn{7}{r}{{Continued on next page}} \\
\caption{Backtesting Strategy Results. We use U for unscaled (no volatility scaling) position sizing, and S for volatility scaled position sizes. }
\label{tab:backtest_results}
\endfoot

\bottomrule \\
\caption{Backtesting Strategy Results. We use U for unscaled (no volatility scaling) position sizing, and S for volatility scaled position sizes. }
\label{tab:backtest_results}
\endlastfoot

buy and hold & 100 & 0.8  & -13.66 & 6.84 & -1.27 & 12.2 \\
 \midrule
Transformer (V) & &  &&&&\\ \midrule

(U) buy-low-sell-high & 94.0 & 0.94  & -10.83 & 8.16 & -1.21 & 14.2\\
(S) buy-low-sell-high &94.0 &0.96  &-17.08 &8.4 &-2.0 &24.1 \\
(U) Momentum &2.0 &0.84  &-0.01 &43.76 &-0.01 &0.148 \\
(S) Momentum &2.0 &0.84  &-0.03 &43.7 &-0.03 &0.307 \\
(U) Mean Reversion   &9.0 &-0.14  &-2.56 &-2.4 &-0.31 &-0.571 \\
(S) Mean Reversion  &9.0 & 0.03 &-5.17 &0.55 &-0.71 &-0.006 \\
 \midrule

Synthesizer  Transformer (R)& &  &&&&\\
 \midrule
 
(U) buy-low-sell-high  &75.0 &1.06  &-7.76 &9.37 &-1.04 &14.1 \\
(S) buy-low-sell-high &75.0 & 1.0 &-12.83 &9.0 &-1.82 &22.9 \\
(U) Momentum &15.0 & 0.94 &-0.93 &10.65 &-0.21 &2.43 \\
(S) Momentum &15.0 &1.09  &-1.95 &12.13 & -0.48&6.58 \\
(U) Mean Reversion   &23.0 & -0.03 &-5.77 &-0.46 &-0.55 &-0.395 \\
(S) Mean Reversion  &23.0 &0.04  &-12.92 &0.55 &-1.31 &-0.361 \\
 \midrule

Synthesizer  Transformer (FR) & &  &&&&\\
 \midrule
 
(U) buy-low-sell-high  & 70.0&0.59  &-13.37 &5.51 &-0.98 & 6.70\\
(S) buy-low-sell-high &70.0 &0.32  &-23.56 &3.12 &-1.77 &5.12 \\
(U) Momentum &17.0 &0.05  &-3.26 &0.72 &-0.28 &0.135 \\
(S) Momentum &17.0 &0.22  &-7.07 &2.8 &-0.64 &1.45 \\
(U) Mean Reversion   &27.0 &0.89  &-4.54 &9.79 &-0.7 &7.68 \\
(S) Mean Reversion  &27.0 &0.84  & -11.76&9.44 &-1.65 &16.8 \\
 \midrule

 Synthesizer Transformer (D)& &  &&&&\\
 \midrule
 
(U) buy-low-sell-high  &78.0 &0.51  &-13.84 &4.65 &-0.98 &5.68 \\
(S) buy-low-sell-high &78.0 &0.41  &-24.03 &3.8 &-1.75 &7.25 \\
(U) Momentum &10.0 &0.81  & -1.5&12.13 &-0.19 &1.88 \\
(S) Momentum &10.0 &0.82  &-3.11 & 12.27&-0.39 &3.97 \\
(U) Mean Reversion   &23.0 &0.79  &-3.73 &9.61 &-0.71 &6.89 \\
(S) Mean Reversion  &23.0 &0.75  &-8.24 &9.22 &-1.48 & 13.2\\
 \midrule

Synthesizer Transformer (FD) r& &  &&&&\\
 \midrule
 
(U) buy-low-sell-high  &71.0 &0.19  &-14.59 & 1.89&-1.01 &1.71 \\
(S) buy-low-sell-high &71.0 &0.06  &-26.18 &0.59 &-1.88 &-0.753 \\
(U) Momentum &13.0 &1.54  &-1.36 &18.78 &-0.23 &4.55 \\
(S) Momentum &13.0 &1.56  &-2.64 &19.07 &-0.47 & 9.60\\
(U) Mean Reversion   &26.0 &1.03  &-3.72 &11.92 &-0.66 &8.67 \\
(S) Mean Reversion  &26.0 &1.03  &-8.39 &12.13 &-1.46 &19.2 \\
 \midrule

Synthesizer  Transformer (DM) & &  &&&&\\
 \midrule
 
(U) buy-low-sell-high  &85.0 &0.76  &-16.11&6.86 &-1.09 &9.96 \\
(S) buy-low-sell-high &85.0 & 0.77 &-25.36&6.95 &-1.78 & 16.21\\
(U) Momentum &4.0 &1.72  &-0.02& 48.36&-0.14 & 3.12\\
(S) Momentum &4.0 &1.72 &-0.05&48.37 &-0.3 &6.73 \\
(U) Mean Reversion  &16.0 &-0.04  &-6.99& -0.74&-0.58 &-0.483 \\
(S) Mean Reversion  &16.0 &0.01  &-15.37&0.2 & -1.32& -0.818 \\
 \midrule

Synthesizer  Transformer (MR) & &  &&&&\\
\midrule

(U) buy-low-sell-high  & 83.0& 0.49 &-16.12 &4.4 &-1.13 &6.17 \\
(S) buy-low-sell-high &83.0 &0.5  &-26.36 &4.5& -1.93&10.0 \\
(U) Momentum &6.0 &1.97  &-0.24 &40.56 &-0.17 &4.32 \\
(S) Momentum &6.0 &1.98  &-0.5 &40.67 &-0.35 & 9.36\\
(U) Mean Reversion   &16.0 &0.07  &-6.97 &1.2 &-0.57 & 0.26\\
(S) Mean Reversion  &16.0 &0.16  &-15.53 &2.95 &-1.35 &1.62 \\

 \end{longtable}}

\begin{figure}[htbp!] 
    \centering
    \begin{subfigure}[t]{.76\linewidth}
    \includegraphics[width=1\textwidth]{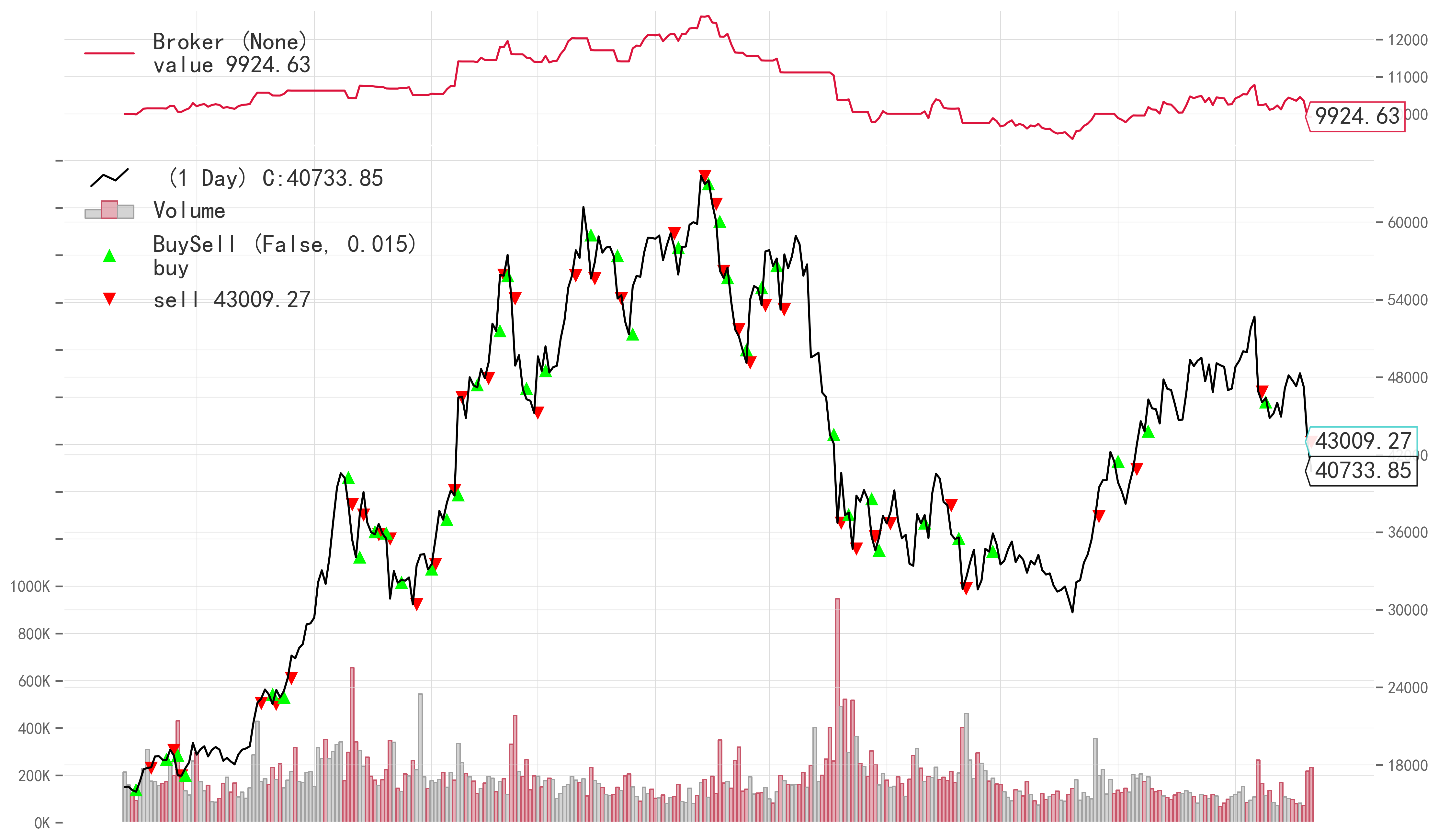}
    \caption{The buy-low-sell-high strategy. }
    \end{subfigure}
      \begin{subfigure}[t]{.76\linewidth}
    \includegraphics[width=1\textwidth]{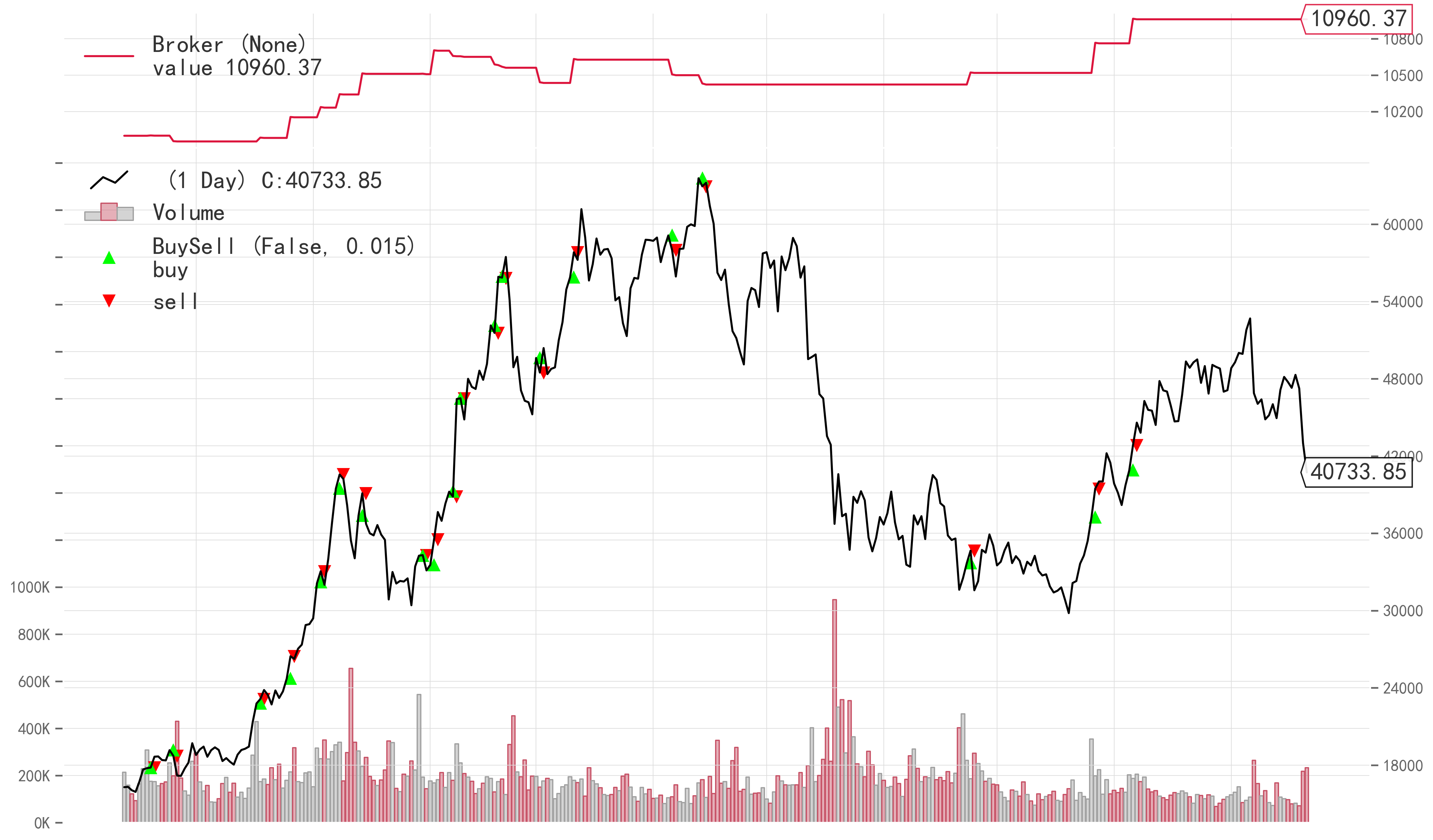}
    \caption{The Momentum strategy. }
    \end{subfigure}
  \begin{subfigure}[t]{.76\linewidth}
    \includegraphics[width=1\textwidth]{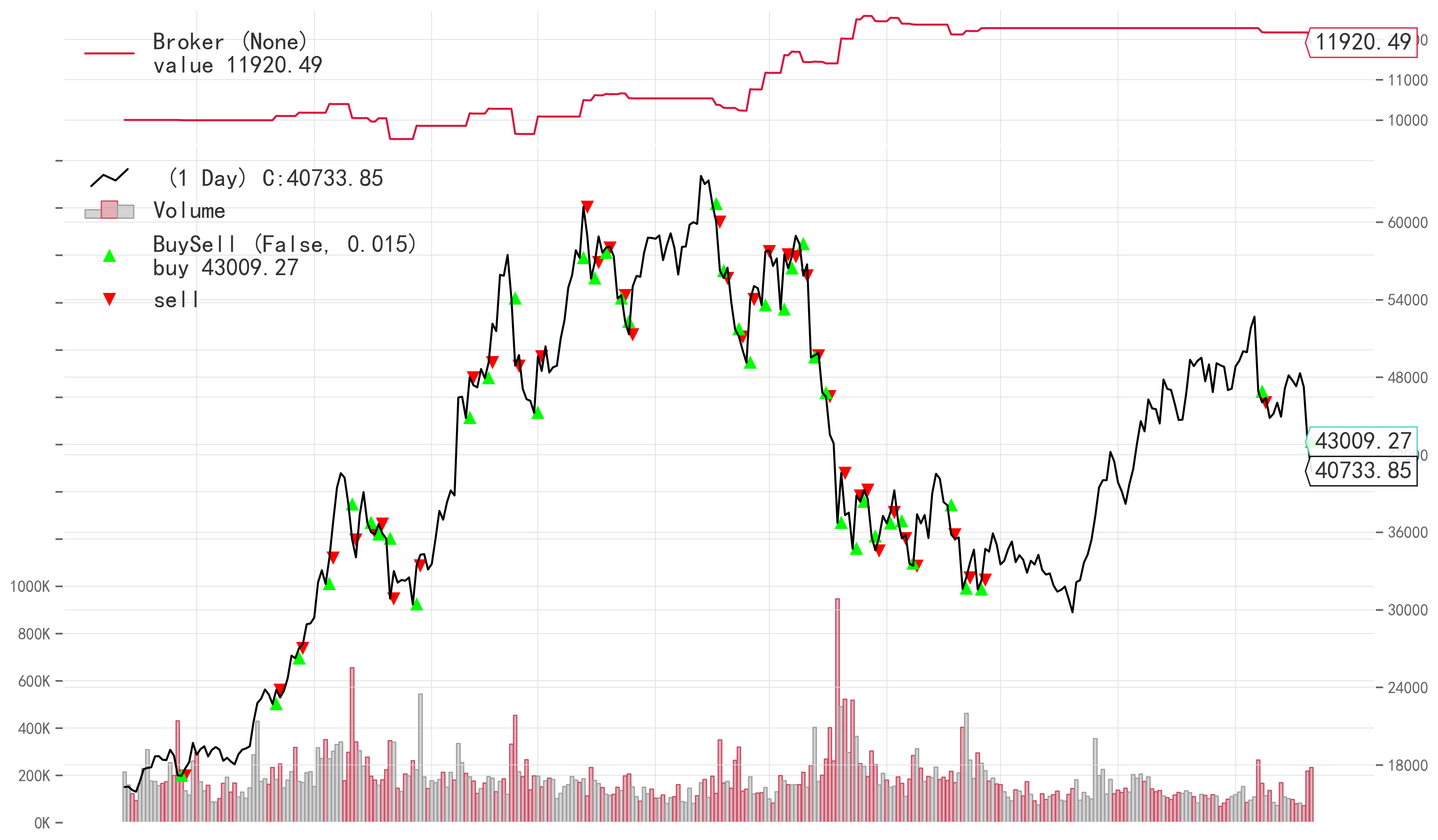}
    \caption{The Mean reversal strategy. }
    \end{subfigure}
    \caption{Backtesting graph of the Synthesizer (FD) model. The red down arrows indicate sell signals, and the green up arrows buy signals, and the red curve at the top shows the evolution of the portfolio value. }
    \label{fig:backtest_fac_dense}
\end{figure}

When looking at one of the better performing models in terms of extreme volatility prediction of the previous section, Synthesizer FD, we notice that the strategies based on this model consistently obtain one of the highest Sharpe ratios. Especially, the momentum and mean reversal strategies (with volatility position scaling), obtain a profit of 9.6\% and 19.2\%. In Figure~\ref{fig:backtest_fac_dense} details are shown of the actual trades for each of the three (scaled) strategies based on the Synthesizer (FD) model. We notice that the most steady increase in total portfolio value is obtained with both the momentum as well as the mean reversal strategy, which is consistent with the results in table. Overall, while these strategies are overly simple and have ample room for improvements, they show the potential of using volatility predictions for risk reduction and finding trading opportunities.


\section{Conclusions}
\label{sec:7}


In this work, we investigate the usefulness of CryptoQuant data (e.g. on-chain analytics, exchange data, miner data) as well as whale-alert tweets for predicting Bitcoin's next-day volatility. The dataset that was analysed in detail, and the correlation between features and next-day volatility was explored. This analysis uncovered the features important for volatility prediction. 

We then propose a deep learning Transformer model to predict extreme volatility spikes. In particular, we developed a Synthesizer Transformer, a state-of-the-art architecture that is known for its computational efficiency due to the elimination of the dot-product attention mechanism. After parameter tuning, we performed detailed experiments wherein we examined the influence of different synthetic attention mechanisms on the model's performance. We also compared the proposed models to baseline models such as LSTM, Vanilla Transformer, and GARCH. The different Synthesizer models outperform all of the baseline models, both in terms of volatility prediction (regression) as well as volatility spike prediction (classification). The proposed Synthesizer Transformer, especially the one with factorised dense attention, manages to obtain state-of-the-art performance when predicting volatility using CryptoQuant data and whale-alert tweets. 

To gain insight into the inner workings of our Transformer model, we used the Captum XAI library. This allowed us to uncover important input features such as `taker buy volume' and `exchange outflow (ma7)', and USDminus (USD flowing out of wallets into exchanges, from whale-alert tweets). We thus confirmed the importance of both on-chain and whale-alert Twitter features for volatility prediction. 

Finally, we integrated our prediction results with several simple baseline trading strategies. The results show that we are able to minimize drawdown while keeping steady profits. Notably, the Synthesizer Transformer with factorized dense attention performs very well and mitigates downside risk while maintaining a steady profit. We also notice that volatility predicted by our models is especially powerful when used to perform volatility scaling of position sizes, as it increases both the PnL as well as the Sharpe ratio. We should note that these strategies are very simple, each with their own strengths and downfalls, and that they should be improved for use in a real scenario, still, even in this simple form, they demonstrate the power and benefits of our volatility prediction model.

In future research, it would be useful to expand the time frame of both the training and test data, to account for more types of markets. It may also be useful to explore this model for other asset types and on different time scales. Currently, our complete model source code (including trained models) is available online\footnote{\url{https://github.com/dorienh/bitcoin_synthesizer}}, so that it may be used by anyone interested in forecasting extreme volatility movements in the Bitcoin market.

\appendix
\include{tables}


 \bibliographystyle{elsarticle-harv} 
 \bibliography{paper}





\end{document}

%% file: tables.tex
\section{Overview of features}

\centering
{\scriptsize \begin{longtable}{lll}
\\ \toprule 
 \textbf{Variable} & \textbf{Description}  \\ \midrule
\endfirsthead

\multicolumn{2}{c}%
{{-- continued from previous page}} \\ 
\toprule 
\textbf{Variable} & \textbf{Description}  \\ \midrule
\endhead

\bottomrule \multicolumn{2}{r}{{Continued on next page}} \\
    \caption{Description of (daily) features used in the proposed framework.  }
    \label{tab:cryptoquant_table} 
\endfoot

\bottomrule \\
    \caption{Description of (daily) features used in the proposed framework.  }
    \label{tab:cryptoquant_table} 
\\ 
\endlastfoot
  
        
        


  
        \textbf{Inter-entity flow:} \\
        etom\_flow\_total & The total amount of BTC transferred from exchanges to mining pools \\
        etom\_transactions\_count\_flow & Number of transactions from exchanges to mining pools\\
        etom\_flow\_mean& Mean amount of BTC transferred from exchanges to mining pools\\
        
        mtoe\_flow\_total& The total amount of BTC transferred from mining pools to exchanges\\
        mtoe\_transactions\_count\_flow& Number of transactions from mining pool to exchange\\
        mtoe\_flow\_mean&Mean amount of BTC transferred from mining pools to exchanges\\
        
        \midrule

        
        \textbf{Exchange flows:} \\
        exchange\_inflow\_total& Total amount of BTC flowing into exchanges\\
        exchange\_inflow\_top10& Total amount of BTC flowing into top 10 exchanges\\
        exchange\_inflow\_mean&  Average daily transaction value for transactions flowing into exchanges\\
        exchange\_inflow\_mean\_ma7& 7-day moving average of mean exchange\_inflow\_mean\\
        
        exchange\_outflow\_total& Total amount of BTC flowing out of exchanges\\
        exchange\_outflow\_top10& Total amount of BTC flowing out of top 10 exchanges\\
        exchange\_outflow\_mean& Average daily transaction value for transactions flowing out of exchanges\\
        exchange\_outflow\_mean\_ma7& 7-day moving average of exchange\_outflow\_mean\_ma7\\
        
        exchange\_addresses\_count\_inflow&  Number of addresses involved in inflow transactions\\
        exchange\_addresses\_count\_outflow&  Number of addresses involved in outflow transactions\\
        
        exchange\_transactions\_count\_inflow& Number of transactions flowing into exchanges\\
        exchange\_transactions\_count\_outflow& Number of transactions flowing out of exchanges\\
        exchange\_minus& Net amount of BTC flowing out of exchanges\\
        exchange\_plus& Net amount of BTC flowing into exchanges\\

        \midrule
        \textbf{Miner flows:}\\
        miner\_inflow\_total& Total amount of BTC  flowing  into mining pool wallets \\
        miner\_inflow\_top10& Total amount of BTC flowing  into top 10 mining pool wallets\\
        miner\_inflow\_mean& Average daily transaction value for transactions flowing into mining\\ & pool wallets\\
        miner\_inflow\_mean\_ma7& 7-day moving average of miner\_inflow\_mean\\
        miner\_outflow\_total& Total amount of BTC flowing  out of mining pool wallets\\
        miner\_outflow\_top10& Total amount of BTC flowing out of  top 10 mining pool wallets\\
        miner\_outflow\_mean& Average daily transaction value for transactions flowing out of mining \\&pool wallets\\
        miner\_outflow\_mean\_ma7& 7-day moving average of miner\_outflow\_mean\\
        miner\_addresses\_count\_inflow&  Number of addresses involved in inflow transactions\\
        miner\_addresses\_count\_outflow&  Number of addresses involved in outflow transactions\\
        miner\_transactions\_count\_inflow& Number of transactions flowing into BTC miner wallets\\
        miner\_transactions\_count\_outflow& Number of transactions flowing out of BTC miner wallets\\
        miner\_minus& Net amount of BTC flowing out of miner wallets\\
        miner\_plus& Net BTC amount of BTC flowing into miner wallets\\

        \midrule
        \textbf{Network indicators:}\\
        cdd& Coins destroyed by flowing into exchanges \\
        sca& The sum of the days of all coins that was in a kept single wallet \\ 

        \midrule
        \textbf{Market data:}\\
        open& Opening price of BTC in USD at the beginning of the day\\
        high& Highest daily price of BTC in USD\\
         low & Lowest daily price of BTC in USD\\
         close& the closing price in USD at the end of the day\\
         volume & Daily amount of BTC traded\\
         open\_interest & The BTC Perpetual Open Interest from derivative exchanges\\
        market\_cap& Total market capitalization of Bitcoin\\
        funding\_rates& Periodic payments to traders based on the difference\\
        
        &between perpetual contract markets and spot prices\\
        
        taker\_buy\_volume&Volume of perpetual swap trades bought by takers\\
        taker\_sell\_volume& Volume of perpetual swap trades sold by takers\\ 
        taker\_buy\_ratio& Ratio of taker\_buy\_volume divided by taker\_total\_volume\\
        taker\_sell\_ratio& Ratio of taker\_sell\_volume divided by taker\_total\_volume\\
        long\_liquidations& Long leveraged positions in BTC that are forced to exit caused by\\& price volatility\\
        short\_liquidations&Short leveraged positions in BTC that are forced to exit caused by\\& price volatility\\
        long\_liquidations\_usd&Total Amount in USD in long leveraged positions that are forced to exit \\ &caused by price volatility\\
        short\_liquidations\_usd&Total Amount in USD in short leveraged positions that are forced to exit \\ &caused by price volatility\\

        \midrule
        \textbf{Market indicator:}\\
        MVRV (Market-Value-to-Realized-Value)& A ratio of market\_cap divided by realized\_cap\\
        
        \midrule
        
 
        \textbf{Flow indicators:} \\
        exchange\_whale\_ratio& Relative size of the top 10 inflows to total inflows of BTC to exchange\\
        fund\_flow\_ratio & Amount of Bitcoin that exchanges own among the amount of Bitcoin sent\\
        & to the blockchain network\\
        MPI (Miners’ Position Index) & An index to understand miners’ behavior by examining \\
        & the total outflow out of miner wallets\\
 
        \midrule 
        
        \textbf{Twitter whale-alerts:}\\
        BTCminus& The amount of Bitcoin flowing out of wallets into exchange\\
        BTCplus& Total amount of Bitcoin flowing into wallets from exchanges\\
        USDminus& Total amount in USD flowing out of wallets into exchanges\\ 
        USDplus& Total amount in USD flowing into wallets from exchanges\\

        \midrule
        \textbf{Technical indicators:}\\
        ema10& 10-day exponential moving average\\
        HL\_sprd& High-low spread.\\
        CO\_sprd& Close-open spread\\
        log\_returns& Logarithmic return of Bitcoin\\


\end{longtable}}

\FloatBarrier

\pagebreak

\section{Correlation of features with volatility}

\centering
{\scriptsize \begin{longtable}{lccc}
\toprule 
\textbf{Feature } & \textbf{R2} & \textbf{Pearson}  & \textbf{Spearman}\\ \midrule
\endfirsthead

\multicolumn{4}{c}%
{{-- continued from previous page}} \\ 
\toprule 
\textbf{Feature } & \textbf{R2} & \textbf{Pearson}  & \textbf{Spearman}\\ \midrule
\endhead

\bottomrule \multicolumn{4}{r}{{Continued on next page}} \\
\caption{Correlation between volatility and different input features. Values greater than 0.1 are marked in bold.  } \label{tab:corr} 
\endfoot

\bottomrule \\
\caption{Correlation between volatility and different input features. Values greater than 0.1 are marked in bold.  } \label{tab:corr}
\\ 
\endlastfoot



 
exchange\_inflow\_total   &\textbf{0.125} & 0.3528 & 0.287 \\
exchange\_outflow\_total  &\textbf{0.102} & 0.3193 & 0.2744 \\

etom\_flow\_total & 0.0014 & 0.0377 & 0.0876  \\
etom\_transactions\_count\_flow & 0.006 & 0.0777 & \textbf{0.1394}  \\

etom\_flow\_mean & 0.0003 & 0.0172 & 0.0126 \\
mtoe\_flow\_total & 0.0035 & 0.0592 & 0.0758 \\
mtoe\_transactions\_count\_flow &0.02  &\textbf{0.1415} & \textbf{0.2341} \\
mtoe\_flow\_mean & 0.0002 & 0.0152 & -0.0961 \\
exchange\_addresses\_count\_inflow &0.0758   &\textbf{0.2752} & \textbf{0.3047} \\
exchange\_addresses\_count\_outflow & 0.0491 &\textbf{0.2217} & \textbf{0.2368} \\
exchange\_inflow\_total &0.1245  &\textbf{0.3528} &\textbf{0.287}\\
exchange\_inflow\_top10 & 0.0589 &\textbf{0.2427} & \textbf{0.2425} \\
exchange\_inflow\_mean &0.0049  &0.0702 & -0.0047 \\
exchange\_inflow\_mean\_ma7 & 0.0007 & -0.0267 & -0.0751 \\
exchange\_transactions\_count\_inflow & 0.0802 &\textbf{0.2833} &  \textbf{0.3042}\\
exchange\_transactions\_count\_outflow & 0.005 & 0.0711 & 0.0499 \\
exchange\_whale\_ratio & 0.042  &\textbf{-0.2048} & \textbf{-0.1863} \\
fund\_flow\_ratio & 0.0429 &\textbf{0.2071} & \textbf{0.1871} \\
mpi & 0.0227 &\textbf{0.1507} & \textbf{0.1716} \\
miner\_addresses\_count\_inflow & 0.0005 & -0.0232& \textbf{0.1058} \\
miner\_addresses\_count\_outflow & 0.0006 & -0.0249 & 0.0227 \\
miner\_inflow\_total & 0.0018 & -0.0428 & -0.026 \\
miner\_inflow\_top10 & 0.0 & -0.0035 & -0.0002 \\
miner\_inflow\_mean & 0.001 & -0.0324 & -0.0455 \\
miner\_inflow\_mean\_ma7 & 0.0034 & -0.0586 & -0.0748 \\
miner\_outflow\_total & 0.0016 & -0.0397 & -0.0185 \\
miner\_outflow\_top10 & 0.0003 & -0.0167 & -0.0181 \\
miner\_outflow\_mean & 0.0011 & -0.0334 & -0.0902 \\
miner\_outflow\_mean\_ma7 & 0.0039 & -0.0622&\textbf{-0.1197}  \\
miner\_transactions\_count\_inflow & 0.0 & 0.0041 & -0.0325 \\
miner\_transactions\_count\_outflow & 0.0007 & 0.0268 & 0.013 \\
market\_cap & 0.0093 & 0.0966 & \textbf{0.2006} \\

long\_liquidations &\textbf{ 0.103}  &\textbf{0.3209} & \textbf{0.1245}\\
short\_liquidations &0.0341  &\textbf{0.1847} &\textbf{0.1193} \\
long\_liquidations\_usd & 0.0537 &\textbf{0.2318} & \textbf{0.1358} \\
short\_liquidations\_usd & 0.0351 &\textbf{0.1874} &\textbf{0.1342}  \\

open & 0.0108 &\textbf{0.104} &  \textbf{0.2085}\\
high & 0.0122 &\textbf{0.1104} &\textbf{ 0.2175} \\
low & 0.0083 & 0.091 & \textbf{0.1962} \\
close & 0.0102  &\textbf{0.1008} & \textbf{0.2058} \\

volume &\textbf{0.1586}  &\textbf{0.3982} &\textbf{ 0.3705} \\
open\_interest & 0.0023 & 0.0483 & 0.0794 \\

mvrv & 0.0177  &\textbf{0.1332} & \textbf{0.1661} \\
cdd & 0.0222  &\textbf{0.1491} &\textbf{0.1868}   \\

sca & 0.0012 & 0.0351 & 0.0816 \\

funding\_rates & 0.0052 & -0.0724 & -0.0057 \\
taker\_buy\_volume & 0.0274  &\textbf{0.1655} &\textbf{ 0.2336} \\
taker\_sell\_volume &0.0274  &\textbf{0.1655} & \textbf{0.2352} \\

taker\_buy\_ratio & 0.0 & 0.0004 & -0.0032 \\
taker\_sell\_ratio & 0.0 & -0.0004 & 0.0032 \\

exchange\_outflow\_total &  0.102 &\textbf{0.3193} &\textbf{0.2744}  \\
exchange\_outflow\_top10 & 0.0538 &\textbf{0.2319} &\textbf{0.2381}  \\
exchange\_outflow\_mean & 0.0473 &\textbf{0.2176} & \textbf{0.206} \\
exchange\_outflow\_mean\_ma7 &0.0253  &\textbf{0.159} &\textbf{0.1773}  \\

HL\_sprd & \textbf{0.3309} &\textbf{0.5752} & \textbf{0.5117} \\
CO\_sprd & 0.0068 & -0.0825 & -0.029 \\

log\_returns & 0.0098 & -0.0992 & -0.0254 \\

ema10 & 0.0121 &\textbf{0.1101} & \textbf{0.2123}  \\
BTCminus & 0.0542 & \textbf{0.2329}  & 0.0649 \\
BTCplus & 0.0 & 0.0058 & 0.0248 \\

USDminus & 0.0151 & \textbf{0.1227}& 0.0671 \\
USDplus & 0.0009 & 0.0293 & 0.035 \\

exchange\_minus & 0.0004 & -0.0197 & -0.0416 \\
exchange\_plus & 0.0454 &\textbf{0.213} &\textbf{0.1104} \\
miner\_minus & 0.0005 & -0.0234 & -0.0076 \\
miner\_plus & 0.0009 & -0.0304 & -0.0292  \\



\end{longtable}}

\section{Descriptive statistics of features}

{\scriptsize \begin{longtable}{lccccccc}
\toprule 
Variable&N&Mean& Std &Minimum& Maximum& Skewness& Kurtosis\\ \midrule
\endfirsthead

\multicolumn{8}{c}%
{{-- continued from previous page}} \\ 
\toprule 
Variable&N&Mean& Std &Minimum& Maximum& Skewness& Kurtosis\\ \midrule
\endhead

\bottomrule \multicolumn{8}{r}{{Continued on next page}} \\
    \caption{Descriptive statistics for the features.}
    \label{tab:before_normalization}
\endfoot

\bottomrule \\
    \caption{Descriptive statistics for the features.}
    \label{tab:before_normalization}
\\ 
\endlastfoot
        etom\_flow\_total &2090 &671.64 &1782.41 &5.31 &40381.38 & 12.24 &199.89 \\
        etom\_transactions\_count\_flow &2090 &324.09 &314.13 &51.00 &5935.00 &8.19 &101.87 \\
        etom\_flow\_mean &2090 &1.81 &2.87 &0.0769 &71.49 &17.19 &374.87 \\
        mtoe\_flow\_total &2090 &924.30 &1757.58 &90.25 &56263.99 &19.77 &535.17 \\
        mtoe\_transactions\_count\_flow &2090 &408.73 &344.20 &46.00 &5602.00 &6.06 &53.43 \\
        mtoe\_flow\_mean &2090 &3.27 &10.28 &0.298 &332.92 &20.48 &562.53 \\
        
        cdd &2090 &1.21e+07 &1.73e+07 &1.33e+06 &3.97e+08 &9.93 &161.97 \\ 
        sca &2090 &1.63e+10 &3.10e+09 &1.14e+10	 &2.23e+10	 &0.389 &-1.18 \\ 
        open &2090 &10961.95 &13905.24 &359.09 &63555.94 &2.08 &3.48 \\ 
        high &2090 &11321.22 &14356.97 &373.97 &64894.67 &2.07 &3.39 \\ 
        low &2090 &10557.51 &13376.04 &352.40 &62016.75 &2.10 &3.57 \\ 
        close &2090 &10981.18 &13918.65 &358.85 &63572.72 &2.08 &3.46  \\
        volume &2090 &104172.70 &97245.49 &3624.27 &1.39e+06 &3.32 &25.21 \\
        open\_interest &907 &3.60e+09 &3.53e+09 &4.38e+08 &1.48e+10 &1.21 &0.178 \\
        market\_cap  &2090 &2.00e+11 &2.62e+11 &5.41e+09 &1.19e+12 &2.10 &3.50 \\
        funding\_rates &1956 &0.00760 &0.0250 &-0.257 & 0.159&0.632 &12.90 \\
        taker\_buy\_volume &2090 &3.48e+09 &5.89e+09 &39306.00&4.71e+10	 &2.51 &7.01 \\
        taker\_sell\_volume  &2090 &3.52e+09 &5.98e+09 &60220.00 &4.83e+10 &2.54 &7.24 \\
        taker\_buy\_ratio &2090 &0.497 &0.0384 &0.266 &0.819 &-0.310 &11.04  \\
        taker\_sell\_ratio &2090 &0.503 &0.0384 &0.181 &0.734 &0.310 &11.04 \\
        long\_liquidations &904 &4212.18 &9965.51 &7.86 &178839.46 & 11.47&178.48 \\
        short\_liquidations &904 &2451.05 &4980.42 &3.47 &113129.98 &13.28 &272.97 \\
        long\_liquidations\_usd &904 &7.07e+07	 &1.43e+08 &89528.66 &2.03e+09	 &6.73 &66.14 \\
        short\_liquidations\_usd &904 &3.89e+07 &5.51e+07 &35850.00 &5.11e+08	 &3.54 &17.95 \\
        MVRV(Market-Value-to-Realized-Value) &2090 &1.90 &0.694 &0.695 &4.84 &0.875 &0.387  \\
        exchange\_whale\_ratio &2090 &0.410 &0.0779 &0.101 &0.734 &0.304 & 0.716 \\
        fund\_flow\_ratio &2090 &0.0669 &0.0327 &0.000811 &0.257 &0.748 &0.884 \\
        MPI(Miners’ Position Index) &2090 &0.0906 &1.28 &-1.79 &17.62 &2.91 &21.27 \\
    
        BTCminus  &2090 & 498.77&2096.25 &0.00 &50983.00 &11.09 &199.47 \\
        BTCplus  &2090 &2110.08 &8208.79 &0.00 &150730.00 &10.95 &152.33 \\
        USDminus  &2090 & 9.33e+06&4.52e+07 &0.00 &1.23e+09	 &13.05 &280.52 \\
        USDplus  &2090 &2.89e+07 &1.40e+08 &0.00 &4.80e+09 &21.58 &671.08 \\
        ema10 &2082 &10924.28 &13792.68 &376.82 &60859.06 &2.08 &3.43  \\
        HL\_sprd &2090 &0.0633 &0.0491 &0.00713 &0.717 &3.14 &21.37 \\
        CO\_sprd  &2090 &0.00301	&0.0410 &-0.392 &0.271 &-0.147 &7.49  \\
        log\_returns  &2090 &0.00217 &0.0414 &-0.493 &0.240	 & -0.843 &12.57 \\
        vol\_future  &2090 &0.561 &0.593 &0.000234 &8.67 &3.34 &25.92 \\   
        
        miner\_inflow\_total &2090 &6365.36 &7371.97 &617.31 &69414.91 &3.56 &14.73 \\
        miner\_inflow\_top10  &2090 &2681.30 &1934.40 &516.95 & 23227.61 &3.48 &19.93 \\
        miner\_inflow\_mean  &2090 &1.64 &0.958 &0.103 &8.71 &2.10 &6.54 \\
        miner\_inflow\_mean\_ma7  &2090 &1.65 &0.813 &0.425 &4.84 &1.55 &2.90 \\
        miner\_outflow\_total  &2090 &6501.13 &8085.25 &751.89 &87732.70 &4.03 &21.22 \\
        miner\_outflow\_top10  &2090 &4329.98 &3799.29 &719.92 &59621.15 &4.99 &42.45 \\
        miner\_outflow\_mean  &2090 &6.39 &4.66 &0.580 &79.36 &4.27 &40.34 \\
        miner\_outflow\_mean\_ma7  &2090 &6.40 &3.29 &1.07 &34.92 & 2.05&10.91 \\
        miner\_addresses\_count\_inflow   &2090 &5323.60 &4411.59 &1664.00 &34392.00 &3.37 &11.17 \\
        miner\_addresses\_count\_outflow   &2090 &4556.38 &5028.56 &672.00 &55463.00 &3.30 &13.13 \\
        miner\_transactions\_count\_inflow   &2090 &3573.62 &2035.90 &1637.00 &21456.00 &3.40 &15.18 \\
        miner\_transactions\_count\_outflow   &2090 &1075.75 &936.26 &282.00 &8833.00 &3.21 &13.83\\
        miner\_minus    &2090 &866.67 &2881.88 &0.00 &36966.66 &6.75 &57.83 \\
        miner\_plus    &2090 &730.33 &2038.84 &0.00 &	32402.41 &7.71 &79.15\\
        
        exchange\_inflow\_total   &2090 &53086.70 &28663.04 &11099.87 &338823.11 &2.25 &11.03  \\
        exchange\_inflow\_top10   &2090 &20998.42 &11238.30 &5053.41 &134045.94 &3.22 &19.60  \\
        exchange\_inflow\_mean   &2090 &1.47 &0.72 &0.405 &7.46 &1.89 &7.05 \\
        exchange\_inflow\_mean\_ma7   &2090 &1.47 &0.602 &0.525 &4.16 &1.16 &0.926 \\
        exchange\_outflow\_total   &2090 &52382.69 &27237.06 &11922.52 & 334359.32&2.17 &11.36 \\
        exchange\_outflow\_top10 &2090 &25372.54 &12905.31 &5475.27 &128484.75 &2.31 &10.34 \\
        exchange\_outflow\_mean  &2090 &4.94 &3.01 &0.435 &29.57 & 1.75&5.97 \\
        exchange\_outflow\_mean\_ma7 &2090 &4.93 &2.55 &1.00 &15.79 &1.03 &0.911 \\
        exchange\_addresses\_count\_inflow &2090 &48372.63 &24352.00 &14052.00 &234312.00 &3.29 & 17.01\\
        exchange\_addresses\_count\_outflow &2090 &43924.36 &34987.83 &8963.00 &397816.00 &3.97 & 23.97\\
        exchange\_transactions\_count\_inflow &2090 &39649.27 &21199.12 &8512.00 &216757.00 &2.94 &15.11 \\
        exchange\_transactions\_count\_outflow &2090 &12393.55 &5948.59 &2478.00 & &2.25 &13.26 \\
        exchange\_minus &2090 &2434.06 &5150.79 &0.00 &108428.54 &6.72 &98.52 \\
        exchange\_plus &2090 &3137.54 &6151.34 &0.00 &89512.88 &5.02 &44.07 \\

\end{longtable}}

{\scriptsize \begin{longtable}{lcccccccc}
\toprule 
Variable&N&Mean& Std &Minimum& Maximum& Skewness& Kurtosis&TR\\ \midrule
\endfirsthead

\multicolumn{9}{c}%
{{-- continued from previous page}} \\ 
\toprule 
Variable&N&Mean& Std &Minimum& Maximum& Skewness& Kurtosis&TR\\ \midrule
\endhead

\bottomrule \multicolumn{9}{r}{{Continued on next page}} \\
    \caption{Descriptive statistics for all variables after postprocessing. The standardization method (TR) used for each value is indicated in the last column with 0: no change - 1: log() - 2: sqrt root - 3: cube root - 4: power of (1/4) - 5: +1 then log().}
    \label{tab:after_normalization}
\endfoot

\bottomrule \\
    \caption{Descriptive statistics for all variables after postprocessing. The standardization method (TR) used for each value is indicated in the last column with 0: no change - 1: log() - 2: sqrt root - 3: cube root - 4: power of (1/4) - 5: +1 then log().}
    \label{tab:after_normalization}
\\ 
\endlastfoot

        etom\_flow\_total &2090 &5.909 &0.935 &1.67 &10.61 &0.514 &2.26 &1 \\
        etom\_transactions\_count\_flow &2090 &5.61 &0.516 &3.93 &8.69 &1.10 &3.73 &1 \\
        etom\_flow\_mean &2090 &0.912 &0.421 &0.07 &4.28 &1.44 &6.12 & 5\\
        mtoe\_flow\_total &2090 &6.51 &0.703 &4.50 &10.9 &0.528 &2.238 & 1\\
        mtoe\_transactions\_count\_flow &2090 &5.84 &0.563 &3.83 &8.63 &0.0680 &2.50 & 1\\
        mtoe\_flow\_mean &2090 &1.14 &0.572 &0.261 &5.81 &2.31 &9.10 & 5\\
        
        cdd & 2090 &15.96 &0.751&14.10 &19.80 &0.719 &0.957 & 1\\
        sca&2090 &23.50 &0.188 &23.16 &23.83 &0.186 &-1.20 & 1\\ 
        open& 2090&8.55 &1.36 &5.88 &11.06 &-0.331 &-0.644 & 1\\ 
        high& 2090&8.58 &1.36 &5.92 &11.08 & -0.338 &-0.637 & 1\\ 
        low&2090 &8.51 &1.35 &5.86 &11.04 &-0.327 &-0.651 & 1\\ 
        close &2090&8.55 &1.36 &5.88 &11.06 &-0.332 &-0.643  & 1\\
        volume &2090 &11.16 &0.967 &8.20 &14.15 &-0.482 &-0.168 & 1\\
        
        open\_interest &2090 &9.35 &10.70 &0.00 &23.42 &0.277 &-1.91 & 1\\
        market\_cap  &2090 &25.11 &1.42 &22.41 &27.80 &-0.344 &-0.645 & 1\\
        funding\_rates &2090 &0.00711 &0.0242 &-0.260 &0.159 &0.706 &13.8 & 0\\
        taker\_buy\_volume &2090 &19.61 &3.40 &10.58 &24.60 &-0.907 &-0.330 & 1\\
        taker\_sell\_volume  &2090 &19.61 &3.38 &11.00 &24.60 &-0.886 &-0.390 & 1\\
        taker\_buy\_ratio &2090 &0.497 &0.0384 &0.266 &0.819 &-0.310 &11.0 & 0\\
        taker\_sell\_ratio &2090 &0.503 &0.0384 &0.181 &0.734 &0.310 &11.0 & 0\\
        
        long\_liquidations &2090 &3.24 &3.81 &0.00 &12.09 &0.421 &-1.64 & 1\\
        short\_liquidations &2090 &3.07 &3.60 &0.00 &11.64 &0.407 &-1.67 & 1\\
        long\_liquidations\_usd &2090 &7.39 &8.52 &0.00 &21.43 &0.312 &-1.86 & 1\\
        short\_liquidations\_usd & 2090&7.23 &8.33 &0.00 &20.05 &0.306 &-1.87 & 1\\
        
        MVRV(Market-Value-to-Realized-Value) &2090 &1.36 &0.244 &0.834 &2.20 &0.459 &-0.153 & 2 \\
        exchange\_whale\_ratio &2090 &0.410 &0.0780 &0.101 &0.734 &0.304 &0.716 & 0\\
        fund\_flow\_ratio &2090 &0.0669 &0.0327 &0.000811 &0.257 &0.748 &0.884 & 0\\
        MPI(Miners’ Position Index) &2090 &0.0906 &1.28 &-1.79 &17.62 &2.91 &21.3 & 0\\
    
        BTCminus  &2090 &1.18 &2.76 &0.00 &10.84 &1.99 &2.11 & 1\\
        BTCplus  &2090&2.53 &3.81 &0.00 &11.92 &0.906 &-1.05 & 1\\
        USDminus  &2090&2.72 &6.27 &0.00 &20.93 &1.89 & 1.63 & 1\\
        USDplus  & 2090&5.47 &8.11 &0.00 &22.30 &0.830 &-1.27 & 1 \\
        ema10 & 2090&8.54 &1.36 &5.93 &11.02 &-0.334 &-0.65 & 0\\
        HL\_sprd & 2090&0.379 &0.0861 &0.192 &0.895 &0.785 &1.203 & 3\\
        CO\_sprd  & 2090&0.00301 &0.0410 &-0.392 &0.271 &-0.147 &7.49 &0 \\
        log\_returns  & 2090&0.00217 &0.0414 &-0.493 &0.249 &-0.843 &12.6 &0 \\
        vol\_future  & 2090&0.786 &0.217 &0.124 &1.72 &-0.00200 &-0.0569 & 4\\   
        
        miner\_inflow\_total &2090 &8.44 &0.690 &6.43 &11.15 &1.18 & 1.79 & 1\\
        miner\_inflow\_top10  &2090 &7.73 &0.530 &6.25 &10.05 &0.759 &0.882 & 1\\
        miner\_inflow\_mean  &2090 &1.14 &0.202 &0.469 &2.06 &0.672 &1.20 & 3\\
        miner\_inflow\_mean\_ma7  &2090 &1.25 &0.294 &0.652 &2.20 &0.786 &0.990 & 2\\
        
        miner\_outflow\_total  & 2090 &8.43 &0.730 &6.62 &11.38 &1.12 &1.51 & 1\\
        miner\_outflow\_top10  & 2090 &8.18 &0.571 &6.58 &11.00 &0.908 &1.35 & 1\\
        miner\_outflow\_mean  & 2090 &1.80 &0.366 &0.834 &4.30 &0.827 &2.50 & 3\\
        miner\_outflow\_mean\_ma7  &2090 &1.55 &0.195 &0..834 &4.30 &0.0365 &0.698 & 1\\
        
        miner\_addresses\_count\_inflow   &2090 &8.42 &0.484 &7.42 &10.45 &1.95 &4.25 & 1\\
        miner\_addresses\_count\_outflow   &2090 &8.13 &0.653 &6.51 &10.92 &1.47 &2.12 & 1\\
        miner\_transactions\_count\_inflow   &2090 &8.09 &0.393 &7.40 &9.97 &1.51 &2.78 & 1\\
        miner\_transactions\_count\_outflow   &2090 &6.76 &0.600 &5.64 &9.09 &1.02 &0.776 & 1\\
        
        miner\_minus    &2090 &2.78 &3.36 &0.00 &10.52 &0.543 &-1.41 & 1\\
        miner\_plus    &2090 &3.62 &3.29 &0.00 &10.39 &-0.0393 &-1.72 & 1\\
        
        exchange\_inflow\_total   &2090 &10.76 &0.493 &9.31 &12.73 &0.00779 &0.294 & 1\\
        exchange\_inflow\_top10   &2090 &9.85 &0.445 &8.53 &11.81 &0.248 &1.08 & 1\\
        exchange\_inflow\_mean    &2090 &1.11 &0.169 &0.740 &1.95 &0.665 &0.719 & 3\\
        exchange\_inflow\_mean\_ma7   &2090 &1.19 &0.234 &0.725 &2.04 &0.782 &-0.0268 & 2\\
        exchange\_outflow\_total   &2090 &10.75 & 0.484 &9.39 &12.72 &-0.0628 &0.238 & 1\\
        exchange\_outflow\_top10 &2090 &10.03 &0.460 &8.61 &11.76 & 0.00670 &0.539 & 1\\
        exchange\_outflow\_mean  &2090 &1.64 &0.321 &0.758 &3.09 &0.400 &0.241 & 3\\
        exchange\_outflow\_mean\_ma7 &2090 &1.45 &0.188 &0.100 &1.99 &0.170 &-0.435 & 1\\
        exchange\_addresses\_count\_inflow &2090 &10.70 &0.398 &9.55 &12.36 &0.608 &1.54 & 1\\
        exchange\_addresses\_count\_outflow &2090 &10.49 &0.616 &9.10 &12.90 &0.172 &0.579 & 1\\
        exchange\_transactions\_count\_inflow &2090 &10.48 &0.461 &9.05 &12.29 &0.0935 &0.770 & 1\\
        exchange\_transactions\_count\_outflow &2090 &9.32 &0.460 &7.82 &11.29 &-0.319 &0.858 & 1\\

        exchange\_minus &2090 &3.70 &4.05 &0.00 &11.59 &0.260 &-1.80 & 1\\
        exchange\_plus &2090 &4.31 &4.13 &0.00 &11.40 &-0.00799 &-1.87 & 1\\

\end{longtable}}

%% file: paper.bbl
\begin{thebibliography}{72}
\expandafter\ifx\csname natexlab\endcsname\relax\def\natexlab#1{#1}\fi
\providecommand{\url}[1]{\texttt{#1}}
\providecommand{\href}[2]{#2}
\providecommand{\path}[1]{#1}
\providecommand{\DOIprefix}{doi:}
\providecommand{\ArXivprefix}{arXiv:}
\providecommand{\URLprefix}{URL: }
\providecommand{\Pubmedprefix}{pmid:}
\providecommand{\doi}[1]{\href{http://dx.doi.org/#1}{\path{#1}}}
\providecommand{\Pubmed}[1]{\href{pmid:#1}{\path{#1}}}
\providecommand{\bibinfo}[2]{#2}
\ifx\xfnm\relax \def\xfnm[#1]{\unskip,\space#1}\fi
\bibitem[{Aharon et~al.(2022)Aharon, Demir, Lau and
  Zaremba}]{aharon2022Twitter}
\bibinfo{author}{Aharon, D.Y.}, \bibinfo{author}{Demir, E.},
  \bibinfo{author}{Lau, C.K.M.}, \bibinfo{author}{Zaremba, A.},
  \bibinfo{year}{2022}.
\newblock \bibinfo{title}{Twitter-based uncertainty and cryptocurrency
  returns}.
\newblock \bibinfo{journal}{Research in International Business and Finance}
  \bibinfo{volume}{59}, \bibinfo{pages}{101546}.
\bibitem[{Akbiyik et~al.(2021)Akbiyik, Erkul, Kaempf, Vasiliauskaite and
  Antulov-Fantulin}]{akbiyik2021ask}
\bibinfo{author}{Akbiyik, M.E.}, \bibinfo{author}{Erkul, M.},
  \bibinfo{author}{Kaempf, K.}, \bibinfo{author}{Vasiliauskaite, V.},
  \bibinfo{author}{Antulov-Fantulin, N.}, \bibinfo{year}{2021}.
\newblock \bibinfo{title}{Ask" who", not" what": Bitcoin volatility forecasting
  with twitter data}.
\newblock \bibinfo{journal}{arXiv preprint arXiv:2110.14317} .
\bibitem[{Akyildirim et~al.(2021)Akyildirim, Goncu and
  Sensoy}]{akyildirim2021prediction}
\bibinfo{author}{Akyildirim, E.}, \bibinfo{author}{Goncu, A.},
  \bibinfo{author}{Sensoy, A.}, \bibinfo{year}{2021}.
\newblock \bibinfo{title}{Prediction of cryptocurrency returns using machine
  learning}.
\newblock \bibinfo{journal}{Annals of Operations Research}
  \bibinfo{volume}{297}, \bibinfo{pages}{3--36}.
\bibitem[{Alessandretti et~al.(2018)Alessandretti, ElBahrawy, Aiello and
  Baronchelli}]{alessandretti2018anticipating}
\bibinfo{author}{Alessandretti, L.}, \bibinfo{author}{ElBahrawy, A.},
  \bibinfo{author}{Aiello, L.M.}, \bibinfo{author}{Baronchelli, A.},
  \bibinfo{year}{2018}.
\newblock \bibinfo{title}{Anticipating cryptocurrency prices using machine
  learning}.
\newblock \bibinfo{journal}{Complexity} \bibinfo{volume}{2018}.
\bibitem[{Bariviera et~al.(2017)Bariviera, Basgall, Hasperu{\'e} and
  Naiouf}]{bariviera2017some}
\bibinfo{author}{Bariviera, A.F.}, \bibinfo{author}{Basgall, M.J.},
  \bibinfo{author}{Hasperu{\'e}, W.}, \bibinfo{author}{Naiouf, M.},
  \bibinfo{year}{2017}.
\newblock \bibinfo{title}{Some stylized facts of the bitcoin market}.
\newblock \bibinfo{journal}{Physica A: Statistical Mechanics and its
  Applications} \bibinfo{volume}{484}, \bibinfo{pages}{82--90}.
\bibitem[{Bergsli et~al.(2022)Bergsli, Lind, Moln{\'a}r and
  Polasik}]{bergsli2022forecasting}
\bibinfo{author}{Bergsli, L.{\O}.}, \bibinfo{author}{Lind, A.F.},
  \bibinfo{author}{Moln{\'a}r, P.}, \bibinfo{author}{Polasik, M.},
  \bibinfo{year}{2022}.
\newblock \bibinfo{title}{Forecasting volatility of bitcoin}.
\newblock \bibinfo{journal}{Research in International Business and Finance}
  \bibinfo{volume}{59}, \bibinfo{pages}{101540}.
\bibitem[{Biswas and Gupta(2019)}]{biswas2019analysis}
\bibinfo{author}{Biswas, B.}, \bibinfo{author}{Gupta, R.},
  \bibinfo{year}{2019}.
\newblock \bibinfo{title}{Analysis of barriers to implement blockchain in
  industry and service sectors}.
\newblock \bibinfo{journal}{Computers \& Industrial Engineering}
  \bibinfo{volume}{136}, \bibinfo{pages}{225--241}.
\bibitem[{Black et~al.(2012)Black, Hashimzade and Myles}]{black2012dictionary}
\bibinfo{author}{Black, J.}, \bibinfo{author}{Hashimzade, N.},
  \bibinfo{author}{Myles, G.}, \bibinfo{year}{2012}.
\newblock \bibinfo{title}{A dictionary of economics}.
\newblock \bibinfo{publisher}{Oxford university press}.
\bibitem[{Blau(2017)}]{blau2017price}
\bibinfo{author}{Blau, B.M.}, \bibinfo{year}{2017}.
\newblock \bibinfo{title}{Price dynamics and speculative trading in bitcoin}.
\newblock \bibinfo{journal}{Research in International Business and Finance}
  \bibinfo{volume}{41}, \bibinfo{pages}{493--499}.
\bibitem[{Bollerslev(1986)}]{bollerslev1986generalized}
\bibinfo{author}{Bollerslev, T.}, \bibinfo{year}{1986}.
\newblock \bibinfo{title}{Generalized autoregressive conditional
  heteroskedasticity}.
\newblock \bibinfo{journal}{Journal of econometrics} \bibinfo{volume}{31},
  \bibinfo{pages}{307--327}.
\bibitem[{Brandvold et~al.(2015)Brandvold, Moln{\'a}r, Vagstad and
  Valstad}]{brandvold2015price}
\bibinfo{author}{Brandvold, M.}, \bibinfo{author}{Moln{\'a}r, P.},
  \bibinfo{author}{Vagstad, K.}, \bibinfo{author}{Valstad, O.C.A.},
  \bibinfo{year}{2015}.
\newblock \bibinfo{title}{Price discovery on bitcoin exchanges}.
\newblock \bibinfo{journal}{Journal of International Financial Markets,
  Institutions and Money} \bibinfo{volume}{36}, \bibinfo{pages}{18--35}.
\bibitem[{Charandabi and Kamyar(2021)}]{charandabi2021prediction}
\bibinfo{author}{Charandabi, S.E.}, \bibinfo{author}{Kamyar, K.},
  \bibinfo{year}{2021}.
\newblock \bibinfo{title}{Prediction of cryptocurrency price index using
  artificial neural networks: a survey of the literature}.
\newblock \bibinfo{journal}{European Journal of Business and Management
  Research} \bibinfo{volume}{6}, \bibinfo{pages}{17--20}.
\bibitem[{Charandabi and Kamyar(2022)}]{charandabi2022survey}
\bibinfo{author}{Charandabi, S.E.}, \bibinfo{author}{Kamyar, K.},
  \bibinfo{year}{2022}.
\newblock \bibinfo{title}{Survey of cryptocurrency volatility prediction
  literature using artificial neural networks}.
\newblock \bibinfo{journal}{Business and Economic Research}
  \bibinfo{volume}{12}.
\bibitem[{Cheah and Fry(2015)}]{cheah2015speculative}
\bibinfo{author}{Cheah, E.T.}, \bibinfo{author}{Fry, J.}, \bibinfo{year}{2015}.
\newblock \bibinfo{title}{Speculative bubbles in bitcoin markets? an empirical
  investigation into the fundamental value of bitcoin}.
\newblock \bibinfo{journal}{Economics letters} \bibinfo{volume}{130},
  \bibinfo{pages}{32--36}.
\bibitem[{Chuan and Herremans(2018)}]{chuan2018modeling}
\bibinfo{author}{Chuan, C.H.}, \bibinfo{author}{Herremans, D.},
  \bibinfo{year}{2018}.
\newblock \bibinfo{title}{Modeling temporal tonal relations in polyphonic music
  through deep networks with a novel image-based representation}, in:
  \bibinfo{booktitle}{Proceedings of the AAAI Conference on Artificial
  Intelligence}.
\bibitem[{Chung et~al.(2014)Chung, Gulcehre, Cho and
  Bengio}]{chung2014empirical}
\bibinfo{author}{Chung, J.}, \bibinfo{author}{Gulcehre, C.},
  \bibinfo{author}{Cho, K.}, \bibinfo{author}{Bengio, Y.},
  \bibinfo{year}{2014}.
\newblock \bibinfo{title}{Empirical evaluation of gated recurrent neural
  networks on sequence modeling}.
\newblock \bibinfo{journal}{arXiv preprint arXiv:1412.3555} .
\bibitem[{Colon et~al.(2021)Colon, Kim, Kim and Kim}]{colon2021effect}
\bibinfo{author}{Colon, F.}, \bibinfo{author}{Kim, C.}, \bibinfo{author}{Kim,
  H.}, \bibinfo{author}{Kim, W.}, \bibinfo{year}{2021}.
\newblock \bibinfo{title}{The effect of political and economic uncertainty on
  the cryptocurrency market}.
\newblock \bibinfo{journal}{Finance Research Letters} \bibinfo{volume}{39},
  \bibinfo{pages}{101621}.
\bibitem[{De~Prado(2018)}]{de2018advances}
\bibinfo{author}{De~Prado, M.L.}, \bibinfo{year}{2018}.
\newblock \bibinfo{title}{Advances in financial machine learning}.
\newblock \bibinfo{publisher}{John Wiley \& Sons}.
\bibitem[{Dimpfl(2017)}]{dimpfl2017bitcoin}
\bibinfo{author}{Dimpfl, T.}, \bibinfo{year}{2017}.
\newblock \bibinfo{title}{Bitcoin market microstructure}.
\newblock \bibinfo{journal}{Available at SSRN 2949807} .
\bibitem[{Ding et~al.(2020)Ding, Wu, Sun, Guo and Guo}]{ding2020hierarchical}
\bibinfo{author}{Ding, Q.}, \bibinfo{author}{Wu, S.}, \bibinfo{author}{Sun,
  H.}, \bibinfo{author}{Guo, J.}, \bibinfo{author}{Guo, J.},
  \bibinfo{year}{2020}.
\newblock \bibinfo{title}{Hierarchical multi-scale gaussian transformer for
  stock movement prediction.}, in: \bibinfo{booktitle}{IJCAI}, pp.
  \bibinfo{pages}{4640--4646}.
\bibitem[{Ding et~al.(2015)Ding, Zhang, Liu and Duan}]{ding2015deep}
\bibinfo{author}{Ding, X.}, \bibinfo{author}{Zhang, Y.}, \bibinfo{author}{Liu,
  T.}, \bibinfo{author}{Duan, J.}, \bibinfo{year}{2015}.
\newblock \bibinfo{title}{Deep learning for event-driven stock prediction}, in:
  \bibinfo{booktitle}{Twenty-fourth international joint conference on
  artificial intelligence}.
\bibitem[{Dyhrberg(2016)}]{dyhrberg2016bitcoin}
\bibinfo{author}{Dyhrberg, A.H.}, \bibinfo{year}{2016}.
\newblock \bibinfo{title}{Bitcoin, gold and the dollar--a garch volatility
  analysis}.
\newblock \bibinfo{journal}{Finance Research Letters} \bibinfo{volume}{16},
  \bibinfo{pages}{85--92}.
\bibitem[{Engle(1982)}]{engle1982autoregressive}
\bibinfo{author}{Engle, R.F.}, \bibinfo{year}{1982}.
\newblock \bibinfo{title}{Autoregressive conditional heteroscedasticity with
  estimates of the variance of united kingdom inflation}.
\newblock \bibinfo{journal}{Econometrica: Journal of the econometric society} ,
  \bibinfo{pages}{987--1007}.
\bibitem[{Fang et~al.(2019)Fang, Bouri, Gupta and Roubaud}]{fang2019does}
\bibinfo{author}{Fang, L.}, \bibinfo{author}{Bouri, E.},
  \bibinfo{author}{Gupta, R.}, \bibinfo{author}{Roubaud, D.},
  \bibinfo{year}{2019}.
\newblock \bibinfo{title}{Does global economic uncertainty matter for the
  volatility and hedging effectiveness of bitcoin?}
\newblock \bibinfo{journal}{International Review of Financial Analysis}
  \bibinfo{volume}{61}, \bibinfo{pages}{29--36}.
\bibitem[{Gkillas et~al.(2021)Gkillas, Tantoula and
  Tzagarakis}]{gkillas2021transaction}
\bibinfo{author}{Gkillas, K.}, \bibinfo{author}{Tantoula, M.},
  \bibinfo{author}{Tzagarakis, M.}, \bibinfo{year}{2021}.
\newblock \bibinfo{title}{Transaction activity and bitcoin realized
  volatility}.
\newblock \bibinfo{journal}{Operations Research Letters} \bibinfo{volume}{49},
  \bibinfo{pages}{715--719}.
\bibitem[{Grinberg(2012)}]{grinberg2012Bitcoin}
\bibinfo{author}{Grinberg, R.}, \bibinfo{year}{2012}.
\newblock \bibinfo{title}{Bitcoin: An innovative alternative digital currency}.
\newblock \bibinfo{journal}{Hastings Sci. \& Tech. LJ} \bibinfo{volume}{4},
  \bibinfo{pages}{159}.
\bibitem[{Hochreiter and Schmidhuber(1997)}]{hochreiter1997long}
\bibinfo{author}{Hochreiter, S.}, \bibinfo{author}{Schmidhuber, J.},
  \bibinfo{year}{1997}.
\newblock \bibinfo{title}{Long short-term memory}.
\newblock \bibinfo{journal}{Neural computation} \bibinfo{volume}{9},
  \bibinfo{pages}{1735--1780}.
\bibitem[{Hoyle and Shephard(2018)}]{hoyle2018volatility}
\bibinfo{author}{Hoyle, E.}, \bibinfo{author}{Shephard, N.},
  \bibinfo{year}{2018}.
\newblock \bibinfo{title}{Volatility scaling's impact on the sharpe ratio}.
\newblock \bibinfo{journal}{Available at SSRN 3279787} .
\bibitem[{Hu et~al.(2021)Hu, Zhao and Khushi}]{hu2021survey}
\bibinfo{author}{Hu, Z.}, \bibinfo{author}{Zhao, Y.}, \bibinfo{author}{Khushi,
  M.}, \bibinfo{year}{2021}.
\newblock \bibinfo{title}{A survey of forex and stock price prediction using
  deep learning}.
\newblock \bibinfo{journal}{Applied System Innovation} \bibinfo{volume}{4},
  \bibinfo{pages}{9}.
\bibitem[{Jagannath et~al.(2021)Jagannath, Barbulescu, Sallam, Elgendi,
  McGrath, Jamalipour, Abdel-Basset and Munasinghe}]{jagannath2021chain}
\bibinfo{author}{Jagannath, N.}, \bibinfo{author}{Barbulescu, T.},
  \bibinfo{author}{Sallam, K.M.}, \bibinfo{author}{Elgendi, I.},
  \bibinfo{author}{McGrath, B.}, \bibinfo{author}{Jamalipour, A.},
  \bibinfo{author}{Abdel-Basset, M.}, \bibinfo{author}{Munasinghe, K.},
  \bibinfo{year}{2021}.
\newblock \bibinfo{title}{An on-chain analysis-based approach to predict
  ethereum prices}.
\newblock \bibinfo{journal}{IEEE Access} \bibinfo{volume}{9},
  \bibinfo{pages}{167972--167989}.
\bibitem[{JAIN(2019)}]{jain2019cryptocurrency}
\bibinfo{author}{JAIN, D.}, \bibinfo{year}{2019}.
\newblock \bibinfo{title}{Cryptocurrency price prediction using transformer: a
  deep learning architecture}.
\newblock Ph.D. thesis.
\bibitem[{Jiang et~al.(2022)Jiang, Zeng, Song and Liu}]{jiang2022forecasting}
\bibinfo{author}{Jiang, K.}, \bibinfo{author}{Zeng, L.}, \bibinfo{author}{Song,
  J.}, \bibinfo{author}{Liu, Y.}, \bibinfo{year}{2022}.
\newblock \bibinfo{title}{Forecasting value-at-risk of cryptocurrencies using
  the time-varying mixture-accelerating generalized autoregressive score
  model}.
\newblock \bibinfo{journal}{Research in International Business and Finance}
  \bibinfo{volume}{61}, \bibinfo{pages}{101634}.
\bibitem[{Jiang(2021)}]{jiang2021applications}
\bibinfo{author}{Jiang, W.}, \bibinfo{year}{2021}.
\newblock \bibinfo{title}{Applications of deep learning in stock market
  prediction: recent progress}.
\newblock \bibinfo{journal}{Expert Systems with Applications}
  \bibinfo{volume}{184}, \bibinfo{pages}{115537}.
\bibitem[{Jung and Choi(2021)}]{jung2021forecasting}
\bibinfo{author}{Jung, G.}, \bibinfo{author}{Choi, S.Y.}, \bibinfo{year}{2021}.
\newblock \bibinfo{title}{Forecasting foreign exchange volatility using deep
  learning autoencoder-lstm techniques}.
\newblock \bibinfo{journal}{Complexity} \bibinfo{volume}{2021}.
\bibitem[{Katsiampa(2017)}]{katsiampa2017volatility}
\bibinfo{author}{Katsiampa, P.}, \bibinfo{year}{2017}.
\newblock \bibinfo{title}{Volatility estimation for bitcoin: A comparison of
  garch models}.
\newblock \bibinfo{journal}{Economics letters} \bibinfo{volume}{158},
  \bibinfo{pages}{3--6}.
\bibitem[{Kharif(2017)}]{kharif2017Bitcoin}
\bibinfo{author}{Kharif, O.}, \bibinfo{year}{2017}.
\newblock \bibinfo{title}{The bitcoin whales: 1,000 people who own 40 percent
  of the market}.
\newblock \bibinfo{journal}{Bloomberg Businessweek} \bibinfo{volume}{8}.
\bibitem[{Khedr et~al.(2021)Khedr, Arif, El-Bannany, Alhashmi and
  Sreedharan}]{khedr2021cryptocurrency}
\bibinfo{author}{Khedr, A.M.}, \bibinfo{author}{Arif, I.},
  \bibinfo{author}{El-Bannany, M.}, \bibinfo{author}{Alhashmi, S.M.},
  \bibinfo{author}{Sreedharan, M.}, \bibinfo{year}{2021}.
\newblock \bibinfo{title}{Cryptocurrency price prediction using traditional
  statistical and machine-learning techniques: A survey}.
\newblock \bibinfo{journal}{Intelligent Systems in Accounting, Finance and
  Management} \bibinfo{volume}{28}, \bibinfo{pages}{3--34}.
\bibitem[{Kim et~al.(2022)Kim, Shin, Choi and Lim}]{kim2022deep}
\bibinfo{author}{Kim, G.}, \bibinfo{author}{Shin, D.H.}, \bibinfo{author}{Choi,
  J.G.}, \bibinfo{author}{Lim, S.}, \bibinfo{year}{2022}.
\newblock \bibinfo{title}{A deep learning-based cryptocurrency price prediction
  model that uses on-chain data}.
\newblock \bibinfo{journal}{IEEE Access} .
\bibitem[{Klein et~al.(2018)Klein, Thu and Walther}]{klein2018Bitcoin}
\bibinfo{author}{Klein, T.}, \bibinfo{author}{Thu, H.P.},
  \bibinfo{author}{Walther, T.}, \bibinfo{year}{2018}.
\newblock \bibinfo{title}{Bitcoin is not the new gold--a comparison of
  volatility, correlation, and portfolio performance}.
\newblock \bibinfo{journal}{International Review of Financial Analysis}
  \bibinfo{volume}{59}, \bibinfo{pages}{105--116}.
\bibitem[{Kokhlikyan et~al.(2020)Kokhlikyan, Miglani, Martin, Wang, Alsallakh,
  Reynolds, Melnikov, Kliushkina, Araya, Yan et~al.}]{kokhlikyan2020captum}
\bibinfo{author}{Kokhlikyan, N.}, \bibinfo{author}{Miglani, V.},
  \bibinfo{author}{Martin, M.}, \bibinfo{author}{Wang, E.},
  \bibinfo{author}{Alsallakh, B.}, \bibinfo{author}{Reynolds, J.},
  \bibinfo{author}{Melnikov, A.}, \bibinfo{author}{Kliushkina, N.},
  \bibinfo{author}{Araya, C.}, \bibinfo{author}{Yan, S.}, et~al.,
  \bibinfo{year}{2020}.
\newblock \bibinfo{title}{Captum: A unified and generic model interpretability
  library for pytorch}.
\newblock \bibinfo{journal}{arXiv preprint arXiv:2009.07896} .
\bibitem[{Lamon et~al.(2017)Lamon, Nielsen and
  Redondo}]{lamon2017cryptocurrency}
\bibinfo{author}{Lamon, C.}, \bibinfo{author}{Nielsen, E.},
  \bibinfo{author}{Redondo, E.}, \bibinfo{year}{2017}.
\newblock \bibinfo{title}{Cryptocurrency price prediction using news and social
  media sentiment}.
\newblock \bibinfo{journal}{SMU Data Sci. Rev} \bibinfo{volume}{1},
  \bibinfo{pages}{1--22}.
\bibitem[{Li et~al.(2019)Li, Liu, Liu, Zhao and Liu}]{li2019neural}
\bibinfo{author}{Li, N.}, \bibinfo{author}{Liu, S.}, \bibinfo{author}{Liu, Y.},
  \bibinfo{author}{Zhao, S.}, \bibinfo{author}{Liu, M.}, \bibinfo{year}{2019}.
\newblock \bibinfo{title}{Neural speech synthesis with transformer network},
  in: \bibinfo{booktitle}{Proceedings of the AAAI Conference on Artificial
  Intelligence}, pp. \bibinfo{pages}{6706--6713}.
\bibitem[{Li et~al.(2002)Li, Ling and McAleer}]{li2002recent}
\bibinfo{author}{Li, W.K.}, \bibinfo{author}{Ling, S.},
  \bibinfo{author}{McAleer, M.}, \bibinfo{year}{2002}.
\newblock \bibinfo{title}{Recent theoretical results for time series models
  with garch errors}.
\newblock \bibinfo{journal}{Journal of Economic Surveys} \bibinfo{volume}{16},
  \bibinfo{pages}{245--269}.
\bibitem[{Liashenko et~al.(2020)Liashenko, Kravets and
  Bobro}]{liashenko2020fractionally}
\bibinfo{author}{Liashenko, O.}, \bibinfo{author}{Kravets, T.},
  \bibinfo{author}{Bobro, O.}, \bibinfo{year}{2020}.
\newblock \bibinfo{title}{Fractionally cointegrated vector autoregression model
  of spread estimation for metals}, in: \bibinfo{booktitle}{2020 10th
  International Conference on Advanced Computer Information Technologies
  (ACIT)}, \bibinfo{organization}{IEEE}. pp. \bibinfo{pages}{643--646}.
\bibitem[{Liu et~al.(2019)Liu, Lin, Liu, Xu, Ren, Diao and
  Yang}]{liu2019Transformer}
\bibinfo{author}{Liu, J.}, \bibinfo{author}{Lin, H.}, \bibinfo{author}{Liu,
  X.}, \bibinfo{author}{Xu, B.}, \bibinfo{author}{Ren, Y.},
  \bibinfo{author}{Diao, Y.}, \bibinfo{author}{Yang, L.}, \bibinfo{year}{2019}.
\newblock \bibinfo{title}{Transformer-based capsule network for stock movement
  prediction}, in: \bibinfo{booktitle}{Proceedings of the First Workshop on
  Financial Technology and Natural Language Processing}, pp.
  \bibinfo{pages}{66--73}.
\bibitem[{Makris et~al.(2021)Makris, Agres and
  Herremans}]{makris2021generating}
\bibinfo{author}{Makris, D.}, \bibinfo{author}{Agres, K.R.},
  \bibinfo{author}{Herremans, D.}, \bibinfo{year}{2021}.
\newblock \bibinfo{title}{Generating lead sheets with affect: A novel
  conditional seq2seq framework}, in: \bibinfo{booktitle}{2021 International
  Joint Conference on Neural Networks (IJCNN)}, \bibinfo{organization}{IEEE}.
  pp. \bibinfo{pages}{1--8}.
\bibitem[{Naimy and Hayek(2018)}]{naimy2018modelling}
\bibinfo{author}{Naimy, V.Y.}, \bibinfo{author}{Hayek, M.R.},
  \bibinfo{year}{2018}.
\newblock \bibinfo{title}{Modelling and predicting the bitcoin volatility using
  garch models}.
\newblock \bibinfo{journal}{International Journal of Mathematical Modelling and
  Numerical Optimisation} \bibinfo{volume}{8}, \bibinfo{pages}{197--215}.
\bibitem[{Nakamoto(2008)}]{nakamoto2008Bitcoin}
\bibinfo{author}{Nakamoto, S.}, \bibinfo{year}{2008}.
\newblock \bibinfo{title}{Bitcoin: A peer-to-peer electronic cash system}.
\newblock \bibinfo{journal}{Decentralized Business Review} ,
  \bibinfo{pages}{21260}.
\bibitem[{Nguyen et~al.(2018)Nguyen, de~Bodisco and Thaver}]{nguyen2018factors}
\bibinfo{author}{Nguyen, T.}, \bibinfo{author}{de~Bodisco, C.},
  \bibinfo{author}{Thaver, R.}, \bibinfo{year}{2018}.
\newblock \bibinfo{title}{Factors affecting bitcoin price in the cryptocurrency
  market: An empirical study.}
\newblock \bibinfo{journal}{International Journal of Business \& Economics
  Perspectives} \bibinfo{volume}{13}.
\bibitem[{Ni et~al.(2008)Ni, Pan and Poteshman}]{ni2008volatility}
\bibinfo{author}{Ni, S.X.}, \bibinfo{author}{Pan, J.},
  \bibinfo{author}{Poteshman, A.M.}, \bibinfo{year}{2008}.
\newblock \bibinfo{title}{Volatility information trading in the option market}.
\newblock \bibinfo{journal}{The Journal of Finance} \bibinfo{volume}{63},
  \bibinfo{pages}{1059--1091}.
\bibitem[{Patel et~al.(2020)Patel, Tanwar, Gupta and Kumar}]{patel2020deep}
\bibinfo{author}{Patel, M.M.}, \bibinfo{author}{Tanwar, S.},
  \bibinfo{author}{Gupta, R.}, \bibinfo{author}{Kumar, N.},
  \bibinfo{year}{2020}.
\newblock \bibinfo{title}{A deep learning-based cryptocurrency price prediction
  scheme for financial institutions}.
\newblock \bibinfo{journal}{Journal of information security and applications}
  \bibinfo{volume}{55}, \bibinfo{pages}{102583}.
\bibitem[{Platanakis and Urquhart(2019)}]{platanakis2019portfolio}
\bibinfo{author}{Platanakis, E.}, \bibinfo{author}{Urquhart, A.},
  \bibinfo{year}{2019}.
\newblock \bibinfo{title}{Portfolio management with cryptocurrencies: The role
  of estimation risk}.
\newblock \bibinfo{journal}{Economics Letters} \bibinfo{volume}{177},
  \bibinfo{pages}{76--80}.
\bibitem[{Radford et~al.(2018)Radford, Narasimhan, Salimans, Sutskever
  et~al.}]{radford2018improving}
\bibinfo{author}{Radford, A.}, \bibinfo{author}{Narasimhan, K.},
  \bibinfo{author}{Salimans, T.}, \bibinfo{author}{Sutskever, I.}, et~al.,
  \bibinfo{year}{2018}.
\newblock \bibinfo{title}{Improving language understanding by generative
  pre-training} .
\bibitem[{Raheman et~al.(2021)Raheman, Kolonin, Goertzel, Hegyk{\"o}zi and
  Ansari}]{raheman2021architecture}
\bibinfo{author}{Raheman, A.}, \bibinfo{author}{Kolonin, A.},
  \bibinfo{author}{Goertzel, B.}, \bibinfo{author}{Hegyk{\"o}zi, G.},
  \bibinfo{author}{Ansari, I.}, \bibinfo{year}{2021}.
\newblock \bibinfo{title}{Architecture of automated crypto-finance agent}, in:
  \bibinfo{booktitle}{2021 International Symposium on Knowledge, Ontology, and
  Theory (KNOTH)}, \bibinfo{organization}{IEEE}. pp. \bibinfo{pages}{10--14}.
\bibitem[{Ramos-P{\'e}rez et~al.(2021)Ramos-P{\'e}rez, Alonso-Gonz{\'a}lez and
  N{\'u}{\~n}ez-Vel{\'a}zquez}]{ramos2021multi}
\bibinfo{author}{Ramos-P{\'e}rez, E.}, \bibinfo{author}{Alonso-Gonz{\'a}lez,
  P.J.}, \bibinfo{author}{N{\'u}{\~n}ez-Vel{\'a}zquez, J.J.},
  \bibinfo{year}{2021}.
\newblock \bibinfo{title}{Multi-transformer: A new neural network-based
  architecture for forecasting s\&p volatility}.
\newblock \bibinfo{journal}{Mathematics} \bibinfo{volume}{9},
  \bibinfo{pages}{1794}.
\bibitem[{Saggu(2022)}]{saggu2022intraday}
\bibinfo{author}{Saggu, A.}, \bibinfo{year}{2022}.
\newblock \bibinfo{title}{The intraday bitcoin response to tether minting and
  burning events: Asymmetry, investor sentiment, and “whale alerts” on
  twitter}.
\newblock \bibinfo{journal}{Finance Research Letters} \bibinfo{volume}{49},
  \bibinfo{pages}{103096}.
\bibitem[{Sapkota(2022)}]{sapkota2022news}
\bibinfo{author}{Sapkota, N.}, \bibinfo{year}{2022}.
\newblock \bibinfo{title}{News-based sentiment and bitcoin volatility}.
\newblock \bibinfo{journal}{International Review of Financial Analysis}
  \bibinfo{volume}{82}, \bibinfo{pages}{102183}.
\bibitem[{Scaillet et~al.(2020)Scaillet, Treccani and
  Trevisan}]{scaillet2020high}
\bibinfo{author}{Scaillet, O.}, \bibinfo{author}{Treccani, A.},
  \bibinfo{author}{Trevisan, C.}, \bibinfo{year}{2020}.
\newblock \bibinfo{title}{High-frequency jump analysis of the bitcoin market}.
\newblock \bibinfo{journal}{Journal of Financial Econometrics}
  \bibinfo{volume}{18}, \bibinfo{pages}{209--232}.
\bibitem[{Shen et~al.(2019)Shen, Urquhart and Wang}]{shen2019does}
\bibinfo{author}{Shen, D.}, \bibinfo{author}{Urquhart, A.},
  \bibinfo{author}{Wang, P.}, \bibinfo{year}{2019}.
\newblock \bibinfo{title}{Does twitter predict bitcoin?}
\newblock \bibinfo{journal}{Economics Letters} \bibinfo{volume}{174},
  \bibinfo{pages}{118--122}.
\bibitem[{Sridhar and Sanagavarapu(2021)}]{sridhar2021multi}
\bibinfo{author}{Sridhar, S.}, \bibinfo{author}{Sanagavarapu, S.},
  \bibinfo{year}{2021}.
\newblock \bibinfo{title}{Multi-head self-attention transformer for dogecoin
  price prediction}, in: \bibinfo{booktitle}{2021 14th International Conference
  on Human System Interaction (HSI)}, \bibinfo{organization}{IEEE}. pp.
  \bibinfo{pages}{1--6}.
\bibitem[{Tay et~al.(2020)Tay, Bahri, Metzler, Juan, Zhao and
  Zheng}]{tay2020Synthesizer}
\bibinfo{author}{Tay, Y.}, \bibinfo{author}{Bahri, D.},
  \bibinfo{author}{Metzler, D.}, \bibinfo{author}{Juan, D.},
  \bibinfo{author}{Zhao, Z.}, \bibinfo{author}{Zheng, C.},
  \bibinfo{year}{2020}.
\newblock \bibinfo{title}{Synthesizer: Rethinking self-attention in transformer
  models. arxiv 2020}.
\newblock \bibinfo{journal}{arXiv preprint arXiv:2005.00743}
  \bibinfo{volume}{2}.
\bibitem[{Tay et~al.(2021)Tay, Bahri, Metzler, Juan, Zhao and
  Zheng}]{tay2021Synthesizer}
\bibinfo{author}{Tay, Y.}, \bibinfo{author}{Bahri, D.},
  \bibinfo{author}{Metzler, D.}, \bibinfo{author}{Juan, D.C.},
  \bibinfo{author}{Zhao, Z.}, \bibinfo{author}{Zheng, C.},
  \bibinfo{year}{2021}.
\newblock \bibinfo{title}{Synthesizer: Rethinking self-attention for
  transformer models}, in: \bibinfo{booktitle}{International Conference on
  Machine Learning}, \bibinfo{organization}{PMLR}. pp.
  \bibinfo{pages}{10183--10192}.
\bibitem[{Thao et~al.(2021)Thao, Balamurali, Roig and
  Herremans}]{thao2021attendaffectnet}
\bibinfo{author}{Thao, H.T.P.}, \bibinfo{author}{Balamurali, B.},
  \bibinfo{author}{Roig, G.}, \bibinfo{author}{Herremans, D.},
  \bibinfo{year}{2021}.
\newblock \bibinfo{title}{Attendaffectnet--emotion prediction of movie viewers
  using multimodal fusion with self-attention}.
\newblock \bibinfo{journal}{Sensors} \bibinfo{volume}{21},
  \bibinfo{pages}{8356}.
\bibitem[{Vaswani et~al.(2017)Vaswani, Shazeer, Parmar, Uszkoreit, Jones,
  Gomez, Kaiser and Polosukhin}]{vaswani2017attention}
\bibinfo{author}{Vaswani, A.}, \bibinfo{author}{Shazeer, N.},
  \bibinfo{author}{Parmar, N.}, \bibinfo{author}{Uszkoreit, J.},
  \bibinfo{author}{Jones, L.}, \bibinfo{author}{Gomez, A.N.},
  \bibinfo{author}{Kaiser, {\L}.}, \bibinfo{author}{Polosukhin, I.},
  \bibinfo{year}{2017}.
\newblock \bibinfo{title}{Attention is all you need}.
\newblock \bibinfo{journal}{Advances in neural information processing systems}
  \bibinfo{volume}{30}.
\bibitem[{Vidal and Kristjanpoller(2020)}]{vidal2020gold}
\bibinfo{author}{Vidal, A.}, \bibinfo{author}{Kristjanpoller, W.},
  \bibinfo{year}{2020}.
\newblock \bibinfo{title}{Gold volatility prediction using a cnn-lstm
  approach}.
\newblock \bibinfo{journal}{Expert Systems with Applications}
  \bibinfo{volume}{157}, \bibinfo{pages}{113481}.
\bibitem[{Vilasuso(2002)}]{vilasuso2002forecasting}
\bibinfo{author}{Vilasuso, J.}, \bibinfo{year}{2002}.
\newblock \bibinfo{title}{Forecasting exchange rate volatility}.
\newblock \bibinfo{journal}{Economics Letters} \bibinfo{volume}{76},
  \bibinfo{pages}{59--64}.
\bibitem[{Wu et~al.(2021)Wu, Tiwari, Gozgor and Leping}]{wu2021does}
\bibinfo{author}{Wu, W.}, \bibinfo{author}{Tiwari, A.K.},
  \bibinfo{author}{Gozgor, G.}, \bibinfo{author}{Leping, H.},
  \bibinfo{year}{2021}.
\newblock \bibinfo{title}{Does economic policy uncertainty affect
  cryptocurrency markets? evidence from twitter-based uncertainty measures}.
\newblock \bibinfo{journal}{Research in International Business and Finance}
  \bibinfo{volume}{58}, \bibinfo{pages}{101478}.
\bibitem[{Yang et~al.(2020)Yang, Ng, Smyth and Dong}]{yang2020html}
\bibinfo{author}{Yang, L.}, \bibinfo{author}{Ng, T.L.J.},
  \bibinfo{author}{Smyth, B.}, \bibinfo{author}{Dong, R.},
  \bibinfo{year}{2020}.
\newblock \bibinfo{title}{Html: Hierarchical transformer-based multi-task
  learning for volatility prediction}, in: \bibinfo{booktitle}{Proceedings of
  The Web Conference 2020}, pp. \bibinfo{pages}{441--451}.
\bibitem[{Yao et~al.(2018)Yao, Yi, Zhai, Lin, Kim, Zhang and
  Lee}]{yao2018predictive}
\bibinfo{author}{Yao, Y.}, \bibinfo{author}{Yi, J.}, \bibinfo{author}{Zhai,
  S.}, \bibinfo{author}{Lin, Y.}, \bibinfo{author}{Kim, T.},
  \bibinfo{author}{Zhang, G.}, \bibinfo{author}{Lee, L.Y.},
  \bibinfo{year}{2018}.
\newblock \bibinfo{title}{Predictive analysis of cryptocurrency price using
  deep learning}.
\newblock \bibinfo{journal}{International Journal of Engineering \& Technology}
  \bibinfo{volume}{7}, \bibinfo{pages}{258--264}.
\bibitem[{Zhang et~al.(2022)Zhang, Qin, Zhang, Bao, Zhang and
  Liu}]{zhang2022Transformer}
\bibinfo{author}{Zhang, Q.}, \bibinfo{author}{Qin, C.}, \bibinfo{author}{Zhang,
  Y.}, \bibinfo{author}{Bao, F.}, \bibinfo{author}{Zhang, C.},
  \bibinfo{author}{Liu, P.}, \bibinfo{year}{2022}.
\newblock \bibinfo{title}{Transformer-based attention network for stock
  movement prediction}.
\newblock \bibinfo{journal}{Expert Systems with Applications}
  \bibinfo{volume}{202}, \bibinfo{pages}{117239}.
\bibitem[{Zheng et~al.(2021)Zheng, Zheng, Wu and Dai}]{zheng2021chain}
\bibinfo{author}{Zheng, P.}, \bibinfo{author}{Zheng, Z.}, \bibinfo{author}{Wu,
  J.}, \bibinfo{author}{Dai, H.N.}, \bibinfo{year}{2021}.
\newblock \bibinfo{title}{On-chain and off-chain blockchain data collection},
  in: \bibinfo{booktitle}{Blockchain Intelligence}.
  \bibinfo{publisher}{Springer}, pp. \bibinfo{pages}{15--39}.
\bibitem[{Zou and Herremans(2022)}]{zou2022multimodal}
\bibinfo{author}{Zou, Y.}, \bibinfo{author}{Herremans, D.},
  \bibinfo{year}{2022}.
\newblock \bibinfo{title}{A multimodal model with twitter finbert embeddings
  for extreme price movement prediction of bitcoin}.
\newblock \bibinfo{journal}{arXiv preprint arXiv:2206.00648} .

\end{thebibliography}
